\documentclass[12pt]{article}

\usepackage{amsmath}
\usepackage{amssymb}
\usepackage{latexsym}
\usepackage{graphicx}
\usepackage{epsfig}

\def\beq{\begin{equation}}
\def\eeq{\end{equation}}
\def\bea{\begin{eqnarray}}
\def\eea{\end{eqnarray}}

\begin{document}

\begin{titlepage}

\vspace*{1cm}
\begin{center}
{\bf \Large Greybody Factors for Scalar Fields emitted by\\[2mm]
a Higher-Dimensional Schwarzschild-de-Sitter\\[4mm] Black-Hole}

\bigskip \bigskip \medskip

{\bf P. Kanti}\footnote{Email: pkanti@cc.uoi.gr},  {\bf T. Pappas}\footnote{Email: thpap@cc.uoi.gr} and
{\bf N. Pappas}\footnote{Email: npappas@cc.uoi.gr}

\bigskip
{\it Division of Theoretical Physics, Department of Physics,\\
University of Ioannina, Ioannina GR-45110, Greece}

\bigskip \medskip
{\bf Abstract}
\end{center}
In this work, we consider the propagation of scalar particles in a
higher-dimensional Schwarzschild-de-Sitter black-hole spacetime, both on the
brane and in the bulk. Our analysis applies for arbitrary partial modes
and for both minimal and non-minimal coupling of the scalar field.
A general expression for the greybody factor is analytically derived
in each case, and its low-energy behaviour is studied in detail.
Its profile in terms of scalar properties (angular-momentum number
and non-minimal coupling parameter) and spacetime properties 
(number of extra dimensions and cosmological constant) is thoroughly
investigated. In contrast to previous studies, the effect of the
cosmological constant is taken into account both close to and far away 
from the black-hole horizon. The dual role of the cosmological
constant, that may act either as a helping agent to the emission of 
scalar particles or as a deterring effect depending on the value
of the non-minimal coupling parameter, is also demonstrated.

\end{titlepage}

\setcounter{page}{1}
%%%%%%%%%%%%%%%%%%%%%%%%%%%%%%%%%%%%%%%%%%%%%%%%%%%%%%%%%%%%%%%%%%%%%%

\section{Introduction}

After the revival of the theories postulating the existence of extra dimensions
at the turn of the previous century \cite{ADD,RS}, the existence and properties
of black-hole solutions in the novel geometrical set-up were widely reconsidered
in the literature. Especially, the effect of the emission of Hawking radiation 
\cite{Hawking} -- being the manifestation of a quantum effect in a gravitational
context -- by higher-dimensional black holes attracted particular attention.
Different types of black-hole spacetimes were considered and different species
of emitted particles were studied. It was soon realised that the Hawking radiation
spectra may provide a wealth of information regarding the higher-dimensional
gravitational background, such as the number of extra spacelike dimensions
\cite{KMR, HK1, graviton-schw}, the magnitude and orientation of the
angular-momentum of the produced black hole \cite{DHKW, CKW, CDKW1, IOP,
CEKT, rot-other, graviton-rot, FST, CDKW3, Stojkovic-ang, Sampaio-ang},
the value of higher-derivative gravitational couplings \cite{GBK} or the
effect of the brane \cite{FS1, tense-brane}
(for a more extensive, but still incomplete,
list of references, see the reviews \cite{Kanti:2004}-\cite{PKEW}).

In the presence of a cosmological constant, the spacetime around such a
black hole takes the form of the higher-dimensional Schwarzschild-de-Sitter
or Tangherlini solution \cite{Tangherlini}. The propagation of a scalar field
in the spacetime around this type of black
hole was studied both analytically and numerically in  \cite{KGB}. However,
the analytical study was limited to the case of the lowest partial mode
and in the very low-energy regime, since the presence of the cosmological
constant increases the complexity of the field equations and makes the
analytic treatment particularly difficult. A later work  \cite{Harmark} extended
the analysis by computing the next-to-leading-order term in the expansion
of the greybody factor in terms of the energy of the scalar particle, focusing
again on the lowest partial mode. More recently, an analysis  \cite{Crispino}
restricted in the 4-dimensional case produced results for the greybody factor
for arbitrary scalar partial modes.

The present work aims at filling a gap in the literature concerning the existence of
analytical results for the greybody factor for arbitrary modes of a scalar field
propagating in the spacetime of a higher-dimensional Schwarzschild-de-Sitter
black hole. The cases of scalar fields living both on the brane and in the bulk
will be considered, as well as the cases of minimal and non-minimal coupling.
An effort will be made to improve the accuracy of our analysis by avoiding
approximations in the form of the metric tensor employed in previous works
 \cite{Harmark, Crispino} that aimed at simplifying the field equations. An
appropriately chosen radial coordinate will allow us to analytically integrate 
the radial equation and take into account the full effect of the bulk cosmological
constant both close to and far away from the black-hole horizon. Analytical
expressions for the greybody factor will be derived, and their approximate
forms up to terms of ${\cal O}(\omega^2)$ for both minimal and non-minimal
couplings will be determined. The asymptotic low-energy values of the greybody
factor will be found in each case, and the role of the non-minimal coupling
parameter will be investigated also in a higher-dimensional context. Finally, the
dual role of the cosmological constant will be examined: in the case of minimal
coupling, this quantity is known \cite{KGB} to enhance the value of the greybody
factor and to give a boost to the emission of scalar fields both in the bulk and
on the brane; in the presence of a non-minimal coupling, the cosmological constant
also acts as an effective mass for the scalar field and thus tends to decrease the
value of the greybody factor. We will show that the desicive factor for the net
effect of the cosmological constant is the value of the non-minimal coupling
parameter. 

The outline of our paper is as follows: in Section 2, we present the 
gravitational background under consideration, both the higher-dimensional
one in the bulk and the induced one on the brane. In Section 3, we focus
on the propagation of scalar fields on the brane: we solve the radial equation
for arbitrary partial modes and value of the non-minimal coupling parameter,
study analytically its low-energy limits and
examine its profile in terms of both particle and spacetime properties.
In Section 4, the whole analysis is performed for a scalar field living in
the bulk. We present our conclusions in Section 5.

%%%%%%%%%%%%%%%%%%%%%%%%%%%%%%%%%%%%%%%%%%%%%%%%%%%%%%

\section{The Gravitational Background}

Let us first consider a purely higher-dimensional gravitational theory 
of the form
%%%%%%%%%%%
\beq
S_D=\int d^{4+n}x\, \sqrt{-G}\,\left(\frac{R_D}{2 \kappa^2_D} - \Lambda\right)\,,
\label{action_D}
\eeq
%%%%%%%%%%%
where $D=4+n$ is the total number of dimensions, $n$ an arbitrary number of
space-like dimensions, $\kappa^2_D=1/M_*^{2+n}$ the higher-dimensional
gravitational constant associated with the fundamental scale of gravity
$M_*$, and $\Lambda$ a positive bulk cosmological constant. 

The variation of the above action with respect to the metric tensor $G_{MN}$
leads to the Einstein's field equations
%%%%%%%%%%%
\beq
R_{MN}-\frac{1}{2}\,G_{MN}\,R_D=-\kappa^2_D G_{MN} \Lambda 
\equiv \kappa^2_D\,T_{MN}\,,
\label{field_eqs}
\eeq
%%%%%%%%%%
where $T_{MN}$ is the bulk energy-momentum tensor. From the above, by contraction
with $G^{MN}$, we may easily find that the higher-dimensional Ricci scalar $R_D$
is given by the following expression
%%%%%%%%%%
\beq
R_D=\frac{2\,(n+4)}{n+2}\,\kappa^2_D \Lambda\,,
\label{RD}
\eeq
%%%%%%%%%%%
in terms of the bulk cosmological constant. 

We will assume the existence of a spherically-symmetric $(4+n)$-dimensional
gravitational background of the form
%%%%%%%%
\beq
ds^2 = - h(r)\,dt^2 + \frac{dr^2}{h(r)} + r^2 d\Omega_{2+n}^2,
\label{bhmetric}
\eeq
%%%%%%%%%%%
where $d\Omega_{2+n}^2$ is the area of the ($2+n$)-dimensional unit sphere
given by
%%%%%%%%
\begin{equation}
d\Omega_{2+n}^2=d\theta^2_{n+1} + \sin^2\theta_{n+1} \,\biggl(d\theta_n^2 +
\sin^2\theta_n\,\Bigl(\,... + \sin^2\theta_2\,(d\theta_1^2 + \sin^2 \theta_1
\,d\varphi^2)\,...\,\Bigr)\biggr)\,,
\label{unit}
\end{equation}
%%%%%%%%%%%
with $0 \leq \varphi < 2 \pi$ and $0 \leq \theta_i \leq \pi$, for $i=1, ..., n+1$. Then,
the field equations (\ref{field_eqs}) lead to the well-known Tangherlini 
solution \cite{Tangherlini}
%%%%%%%%%%%
\beq
h(r) = 1-\frac{\mu}{r^{n+1}} - \frac{2 \kappa_D^2\,\Lambda r^2}{(n+3) (n+2)}\,,
\label{h-fun}
\eeq
%%%%%%%%%%
describing a higher-dimensional Schwarzschild-de-Sitter spacetime. In the above,
the parameter $\mu$ is given by \cite{MP}
%%%%%%%%%
\beq
\mu=\frac{\kappa^2_D M}{(n+2)}\,\frac{\Gamma[(n+3)/2]}{\pi^{(n+3)/2}}\,,
\eeq
%%%%%%%%%%
where $M$ is the black-hole mass. The equation $h(r)=0$ has, in principle,
$(n+3)$ roots, however, not all of them are real and positive. Depending
on the values of the parameters $M$ and $\Lambda$, the Schwarzschild-de-Sitter
spacetime, either four or higher-dimensional, may have two, one or zero
horizons \cite{Molina}. In the context of the present analysis, the values
of our parameters will be chosen so that the spacetime always has two horizons,
the black-hole one $r_h$ and the cosmological one $r_c$, with the region of
interest being the area in between ($r_h<r<r_c$).

According to the brane-world models \cite{ADD,RS}, four-dimensional observers
are restricted to live on the brane. Therefore, although the gravitational
background is generically higher-dimensional, all phenomenologically 
interesting effects take place in the projected-on-the-brane spacetime.
This follows by fixing the value of the extra angular coordinates,
$\theta_i=\pi/2$, for $i=2, ..., n+1$, and has the form 
%%%%%%%%
\beq
ds^2 = - h(r)\,dt^2 + \frac{dr^2}{h(r)} + r^2\,(d\theta^2 + \sin^2\theta\,
d\varphi^2)\,.
\label{metric_brane}
\eeq
%%%%%%%%
Note that the metric function $h(r)$ is still given by Eq. (\ref{h-fun}).
The above spacetime is neither a vacuum solution nor a solution describing 
a 4-dimensional Schwarzschild-de-Sitter spacetime. Instead, it satisfies
the projected-on-the-brane Einstein's field equations in the presence of
an effective energy-momentum tensor. By explicit calculation, one may
find that the curvature of this 4-dimensional background is given by
the expression
%%%%%%%%%%%%%
\beq
R_4=\frac{24 \kappa_D^2 \Lambda}{(n+2) (n+3)} + \frac{n(n-1)\mu}{r^{n+3}}\,,
\label{R4}
\eeq
%%%%%%%%%
and it is clearly generated by a non-trivial brane distribution of matter
according to the AdS/CFT correspondence.

In what follows, we will assume that the above higher-dimensional black
hole emits scalar particles both in the bulk and on the brane. These
particles will carry small quanta of mass or energy compared to the
black hole mass so that the gravitational background, both bulk and brane,
will remain unchanged. 

%%%%%%%%%%%%%%%%%%%%%%%%%%%%%%%%%%%%%%%%%%%%%%%%%%%%%%%%%%%%%%%%%%%%%%%%%%%

\section{Emission of Scalar Particles on the Brane} 

We will first focus on the case of the emission of brane-localised scalar
particles as the most phenomenologically interesting one. We will consider
a general scalar field theory in which the scalar field may couple to 
gravity either minimally or non-minimally, namely
%%%%%%%%%%%%%%%
\beq
S_4=\frac{1}{2}\int d^4x \,\sqrt{-g}\left[\left(1-\xi \Phi^2\right) R_4
- \,\partial_\mu \Phi\,\partial^\mu \Phi \right]\,.
\label{action4}
\eeq
%%%%%%%%%%%%%%
In the above expression, $g_{\mu\nu}$ is the projected metric tensor on the
brane and $R_4$ the corresponding brane curvature given in Eqs. (\ref{metric_brane})
and (\ref{R4}), respectively. Also, $\xi$ is an arbitrary constant parameter
that determines the magnitude of the coupling between the scalar and the
gravitational field on the brane with the value $\xi=0$ corresponding to
the minimal coupling.  

The equation of motion of the aforementioned scalar field has the form
%%%%%%%%%%%%
\beq
\frac{1}{\sqrt{-g}}\,\partial_\mu\left(\sqrt{-g}\,g^{\mu\nu}\partial_\nu \Phi\right)
=\xi R_4\,\Phi\,. 
\label{field-eq-brane}
\eeq
%%%%%%%%%%%%%
Assuming the factorized ansatz
%%%%%%%%%
\begin{equation}
\Phi(t,r,\theta,\varphi)= e^{-i\omega t}\,R(r)\,Y(\theta,\varphi)\,,
\label{facto}
\end{equation}
%%%%%%%%%
where $Y(\theta,\varphi)$ are the scalar spherical harmonics, and using the
well-known eigenvalue equation that these satisfy, we obtain the following
decoupled radial equation for the function $R(r)$
%%%%%%%%%%
\beq
\frac{1}{r^2}\,\frac{d \,}{dr} \biggl(hr^2\,\frac{d R}{dr}\,\biggr) +
\biggl[\frac{\omega^2}{h} -\frac{l(l+1)}{r^2}-\xi R_4\biggr] R=0\,.
\label{radial_brane}
\eeq
%%%%%%%%%%%

%%%%%%%%%%%%%%%%%%%%%
\begin{figure}[t]
  \begin{center}
\mbox{\includegraphics[width = 0.51 \textwidth] {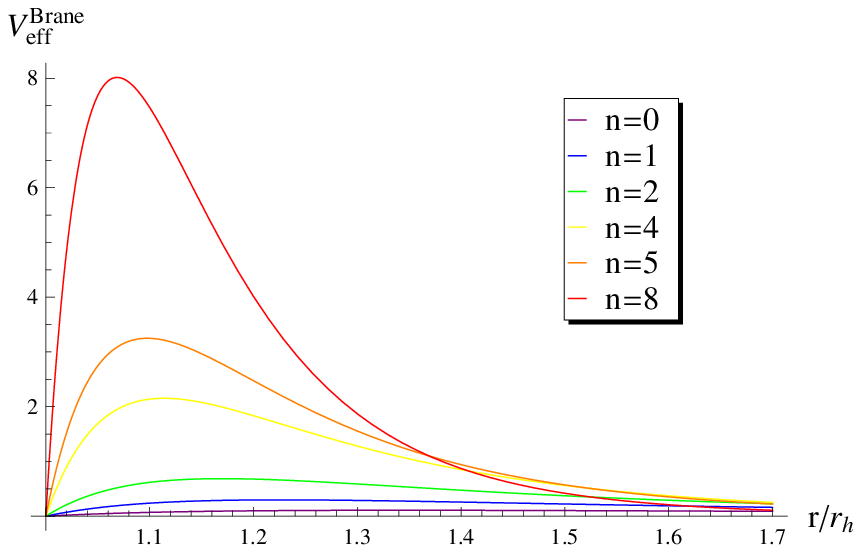}}
\hspace*{-0.6cm} {\includegraphics[width = 0.51 \textwidth]
{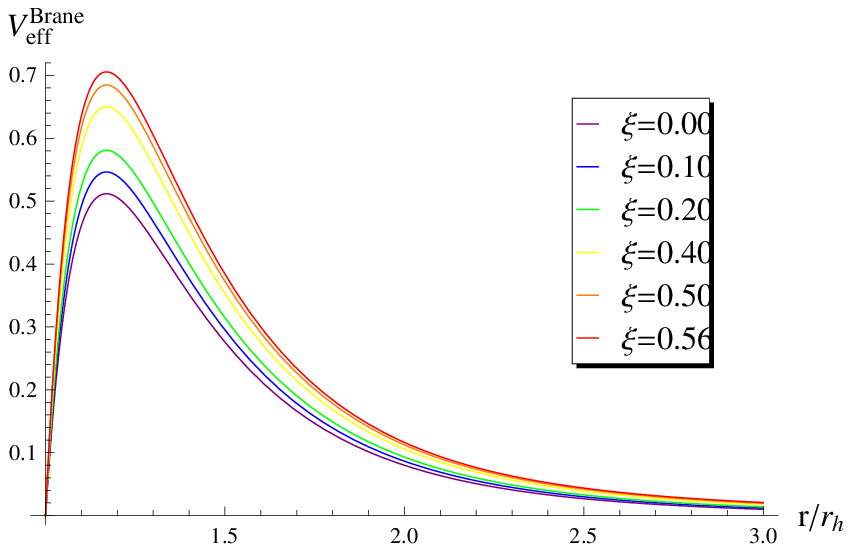}}
    \caption{Effective potential for brane scalar fields for: \textbf{(a)}
	$l=0$, $\Lambda=0.01\,r_h^{-2}$, $\xi=0.5$ and variable $n=0,1,2,4,5,8$ (from bottom
	to the top), and \textbf{(b)} $l=0$, $\Lambda=0.01\,r_h^{-2}$, $n=2$ and variable
	$\xi=0,0.1,0.2,0.4,0.5,0.56$ (again, from bottom to top).}
   \label{pot_brane_nxi}
  \end{center}
\end{figure}
%%%%%%%%%%%%%%%%%

Before attempting to analytically solve the above equation in order to
determine the greybody factor for the emission of brane-localised scalar
fields by a higher-dimensional Schwarzschild-de-Sitter black hole, let
us first study the profile of the effective potential that characterizes
such an emission process. We redefine the radial function as $u(r)=r R(r)$
and use the tortoise coordinate $dr_*=dr/h(r)$; then, Eq. (\ref{radial_brane})
may be re-written in the Schr\"odinger-like form
%%%%%%%%%%%%
\beq
-\frac{d^2 u(r)}{dr^2_*}+ V_{\rm eff}^{\rm brane}\,u(r)=\omega^2 \,u(r)\,,
\label{Schrodinger}
\eeq
%%%%%%%%%%%%
where the effective potential has the form
%%%%%%%%%%%%
\beq
V_{\rm eff}^{\rm brane}=h(r) \left[\frac{l(l+1)}{r^2}+\xi R_4 +
\frac{1}{r}\,\frac{dh}{dr}\right]\,,
\eeq
%%%%%%%%%%%%
or more explicitly
%%%%%%%%%%%%
\beq
V_{\rm eff}^{\rm brane}=h(r) \left\{\frac{l(l+1)}{r^2}+
\frac{4\kappa^2_D \Lambda(6\xi-1)}{(n+2)(n+3)}\,
 + \frac{\mu}{r^{n+3}}\left[\,(n+1)+\xi n(n-1)\,\right]\right\}\,.
 \label{pot_brane}
\eeq
%%%%%%%%%%%%
The effective potential vanishes at the two horizons, $r_h$ and $r_c$, due
to the vanishing of the metric function $h(r)$.
The parameter $\mu$ can be eliminated by using the equation for
the black-hole horizon, i.e.
%%%%%%%%%%%
\beq
\mu=r_h^{n+1}\left(1-\frac{2\kappa^2_D \Lambda r_h^2}{(n+2)(n+3)}\right),
\label{mu}
\eeq
%%%%%%%%%%
and, then, by fixing the value of the black-hole horizon, 
one can study the profile of the effective potential in terms of the 
angular-momentum number $l$, number of extra dimensions $n$, coupling
constant $\xi$ and cosmological constant $\Lambda$.

%%%%%%%%%%%%%%%%%%%%%
\begin{figure}[t]
  \begin{center}
  \begin{tabular}{c}
\mbox{\includegraphics[width = 0.48 \textwidth] {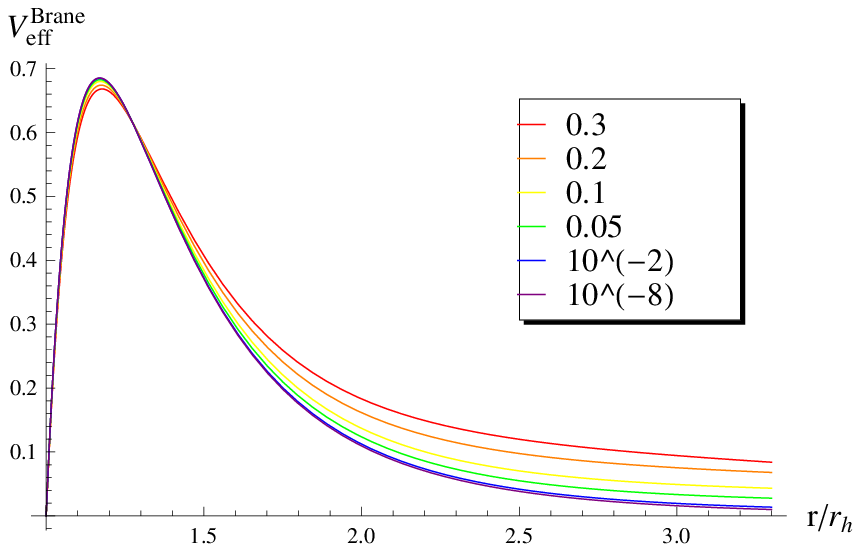}}
\hspace*{-0.1cm} {\includegraphics[width = 0.48 \textwidth]
{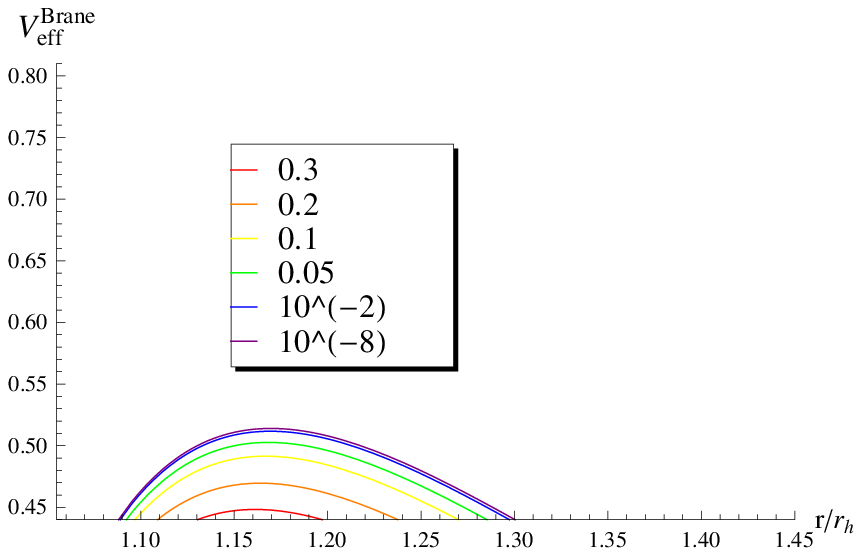}}\\
\mbox{\includegraphics[width = 0.49 \textwidth] {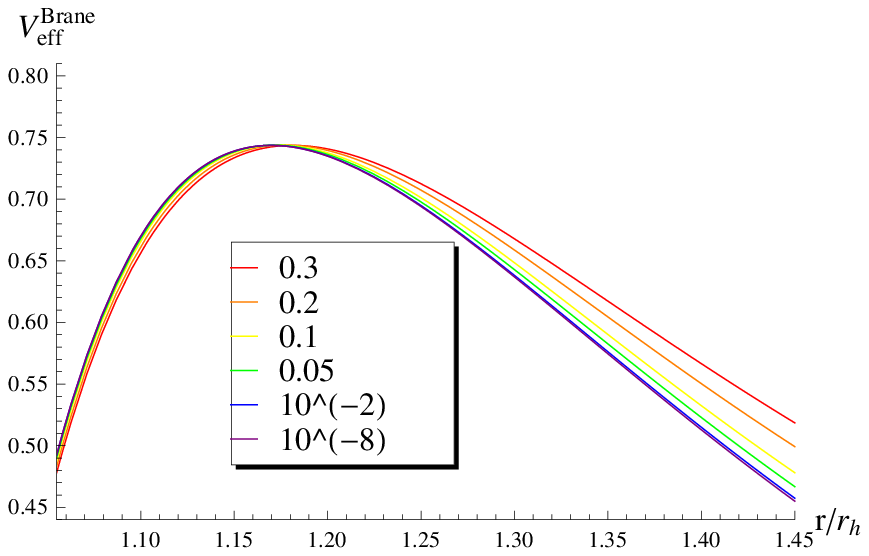}}
\hspace*{-0.2cm} {\includegraphics[width = 0.49 \textwidth]
{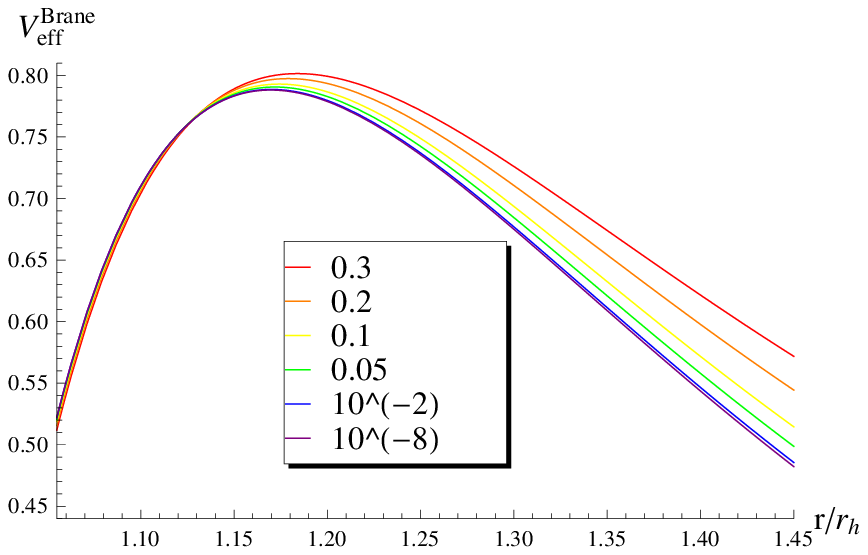}} \end{tabular}
    \caption{Effective potential for brane scalar fields for $l=0$, $n=2$,
    and variable $\Lambda=10^{-8}, 10^{-2},0.05,0.1,0.2,0.3$ (in units of $r_h^{-2}$),
    and \textbf{(a)} wide view for $\xi=0.5$, and magnifications of the
    peak area for \textbf{(b)} $\xi=0$,  \textbf{(c)} $\xi=0.67$ and 
    \textbf{(d)} $\xi=0.8$.}
   \label{pot_brane_Lam}
  \end{center}
\end{figure}
%%%%%%%%%%%%%%%%%

From Eq. (\ref{pot_brane}), it is obvious that the height of the gravitational
barrier increases with the angular-momentum number $l$ leading to the 
well-known suppression of the emission of partial modes with large $l$; however,
the dependence on the other three parameters of the model is less clear.
Therefore, in Fig. \ref{pot_brane_nxi}, we depict the profile of the effective
potential in terms of the number of extra dimensions $n$ and coupling
constant $\xi$: the increase in the value of $n$ causes again an increase
in the height of the gravitational barrier and subsequently the suppression
of the emission of scalar fields on the brane as it was shown in \cite{KGB};
a similar effect arises with the increase of $\xi$ which was also noted
in the context of a purely 4-dimensional theory \cite{Crispino}. In
Fig. \ref{pot_brane_Lam}, we display the dependence of the barrier
on the cosmological constant. The dependence here is more subtle, as
the magnifications of the peak area, shown in Figs. \ref{pot_brane_Lam}(b,c,d),
demonstrate:
for a minimally-coupled scalar field, i.e. $\xi=0$, the barrier
decreases with $\Lambda$ and the emission of scalar particles is
thus enhanced, as it was demonstrated in \cite{KGB}; as the value
of the coupling constant increases, the effect of $\Lambda$ on the
height of the barrier is gradually reduced; finally, beyond a critical
value of $\xi$, the situation is reversed with the barrier now
increasing with any increase in the value of the cosmological constant.

%%%%%%%%%%%%%%%%%%%%%%%%%%%%%%%%%%%

\subsection{The Analytical Solution} 

The differential equation (\ref{radial_brane}) cannot be analytically
solved over the whole radial regime, even in the absence of the
non-minimal coupling. Therefore, an approximate method must be applied
in which the radial equation is solved at specific radial regimes,
i.e. close to the black-hole and cosmological horizons, and then the
corresponding solutions are smoothly matched at the low-energy limit.
For a Schwarzschild-de-Sitter spacetime, this technique was applied 
for the lowest partial mode $l=0$ in the context of a higher-dimensional
model \cite{KGB, Harmark} for minimally-coupled scalar fields, and more
recently for arbitrary $l$ and non-vanishing $\xi$ in the purely
4-dimensional case \cite{Crispino}. We will now address the most general
case  of a scalar field propagating on the brane background 
(\ref{metric_brane}) and with or without a non-minimal coupling. 

We will start by solving the radial equation in the regime close to the
black-hole horizon. Previous studies \cite{Harmark, Crispino}, in 
order to simplify the analysis, have ignored the presence of the
cosmological constant close to the black-hole area and approximated
the metric by a purely Schwarzschild one, either four or higher-dimensional.
In contrast to this, and in an attempt to make our results as accurate as
possible, we will keep all effects of the presence of the cosmological
constant to the black-hole physics. But then, finding the appropriate radial
coordinate transformation, necessary to bring Eq. (\ref{radial_brane}) 
to a more familiar form, turns out to be a real challenge. To our
knowledge, the most convenient such transformation is the following
%%%%%%%%%%
\beq
r \rightarrow f(r) = \frac{h(r)}{1- \tilde \Lambda r^2}\,,
\label{newco-f}
\eeq
%%%%%%%%%%
where, for convenience, we have defined 
$\tilde \Lambda \equiv 2 \kappa_D^2 \Lambda/(n+2)(n+3)$. Also, henceforth,
we will set $\kappa^2_D=1$. The new variable $f$ may be alternatively
written as
%%%%%%%%%
\beq
f(r)=1-\frac{\mu}{r^{n+1}}\,\frac{1}{1-\tilde \Lambda r^2}=
1-\biggl(\frac{r_h}{r}\biggr)^{n+1}\,\frac{(1-\tilde \Lambda r_h^2)}
{(1-\tilde \Lambda r^2)}\,,
\eeq
%%%%%%%%%
and thus ranges from 0, at $r \simeq r_h$, to 1 as $r \gg r_h$. In addition, 
it satisfies the relation
%%%%%%%%%%
\beq
\frac{df}{dr}=\frac{1-f}{r}\,\frac{A(r)}{1-\tilde \Lambda r^2}\,,
\eeq
%%%%%%%%%%
where $A(r) \equiv (n+1)-(n+3) \tilde\Lambda r^2$. By using the above,
Eq. (\ref{radial_brane}) can be re-written, at $r \simeq r_h$, as:
%%%%%%%%%%%
\begin{eqnarray}
&~& \hspace*{-1.6cm} 
f\,(1-f)\,\frac{d^2 R}{df^2} + (1-B_h\,f)\,\frac{d R}{df}
+ \biggl[\,\frac{(\omega r_h)^2}{A_h^2 f} 
- \frac{\lambda_h\,(1-\tilde \Lambda r_h^2)}
{A_h^2 (1-f)}\,\biggr]R=0\,.
\label{NH-1}
\end{eqnarray}
%%%%%%%%%%
In the above, $A_h$ is the value of $A(r)$ evaluated at $r=r_h$, and we have
further defined
%%%%%%%%%
\beq
B_h \equiv 1+ \frac{n}{A_h}\,(1-\tilde \Lambda r_h^2) +
\frac{4\tilde \Lambda r_h^2}{A_h^2}\,, \qquad \lambda_h=l(l+1)+\xi R^{(h)}_4 r_h^2\,,
\eeq
%%%%%%%%%%%
with $R^{(h)}_4$ the brane curvature (\ref{R4}) evaluated at $r=r_h$.

We now make the following field redefinition: $R(f)=f^{\alpha_1} (1-f)^{\beta_1} F(f)$.
Then, Eq. (\ref{NH-1}) takes the form of a hypergeometric equation,
%%%%%%%
\beq
f\,(1-f)\,\frac{d^2 F}{df^2} + [c_1-(1+a_1+b_1)\,f]\,\frac{d F}{df} -a_1b_1\,F=0\,,
\label{hyper}
\eeq
%%%%%%%%%
under the identifications
%%%%%%%%%
\begin{eqnarray}
a_1=\alpha_1 + \beta_1 +B_h-1\, \qquad 
b_1=\alpha_1 + \beta_1\,, \qquad c_1=1+ 2 \alpha_1\,.
\label{abc_bh}
\end{eqnarray}
%%%%%%%%%%
We then need to determine the power coefficients $\alpha_1$ and $\beta_1$; these
are found by solving the following second-order algebraic equations:
%%%%%%%%%
\beq
\alpha_1^2 +\frac{\omega^2 r_h^2}{A_h^2}=0\,,
\label{alpha-eq}
\eeq
%%%%%%%%%
and
%%%%%%%%%
\beq
\beta_1^2 + \beta_1\,(B_h-2) + \frac{\omega^2 r_h^2}{A_h^2} 
- \frac{\lambda_h\,(1-\tilde \Lambda r_h^2)}{A_h^2}=0\,. \label{beta-eq}
\eeq
%%%%%%%%%
The corresponding solutions are
%%%%%%%%%%%%%%
\begin{equation}
\alpha_1^{\,(\pm)} = \pm \frac{i \omega r_h}{A_h}\,,
\label{sol-a}
\end{equation}
%%%%%%%%%%
and 
%%%%%%%%%%%%
\begin{equation}
\beta_1^{\,(\pm)} =\frac{1}{2}\,\biggl[\,(2-B_h)\pm \sqrt{(B_h-2)^2 + 
\frac{4\lambda_h\,(1-\tilde\Lambda r_h^2)}{A_h^2}}\,\,\biggr]\,.
\label{sol-be}
\end{equation}
%%%%%%%%%%%%%

The general solution of the hypergeometric equation (\ref{hyper}), combined
with the relation between $R(f)$ and $F(f)$, leads to the following 
expression for the radial function $R(f)$ in the near-black-hole-horizon regime:
%%%%%%%%
\begin{eqnarray}
&& \hspace*{-1cm}R_{BH}(f)=A_1 f^{\alpha_1}\,(1-f)^{\beta_1}\,F(a_1,b_1,c_1;f)
\nonumber \\[1mm] && \hspace*{2cm} +\,
A_2\,f^{-\alpha_1}\,(1-f)^{\beta_1}\,F(a_1-c_1+1,b_1-c_1+1,2-c_1;f)\,,
\label{BH-gen}
\end{eqnarray}
%%%%%%%%%
\noindent
where $A_{1,2}$ are arbitrary constants. Near the horizon, $f$ goes to
zero, and the above reduces to
%%%%%%%%%%
\beq
R_{BH}(f) \simeq A_1\,f^{\alpha_1} + A_2\,f^{-\alpha_1}\,.
\label{BH-near-1}
\end{equation}
%%%%%%
The two choices therefore for the parameter $\alpha_1$ are equivalent, under
the simultaneous interchange of the arbitrary coefficients $A_{1,2}$. 
As we observed earlier, the effective potential vanishes at the black-hole
horizon, thus we expect the general solution there to have the form
of free plane waves, i.e.
%%%%%%%%%%%%%
\beq
R_{BH}(r_*)=\frac{u_{BH}(r_*)}{r_h}=\tilde A_1 \,e^{-i\omega r_*} +
\tilde A_2 \,e^{i\omega r_*}\,. \label{BH-near-2}
\eeq
%%%%%%%%%%%%%
Clearly, for $f \propto e^{A_h r_*/r_h}$, the two expressions match
under the proper redefinitions of the integration constants. As a result, the 
two choices $\alpha_1^{\,(\pm)}$ simply interchange the incoming and
outgoing plane waves at the black-hole horizon. To remove this
arbitrariness, we set $\alpha_1=\alpha_1^{\,(-)}$. Then, imposing the
boundary condition that no outgoing waves are to be found just outside the
black-hole horizon, demands setting $A_2=\tilde A_2=0$.
Finally, the sign in the expression of the $\beta_1$ coefficient may be
decided from the criterion for the convergence of the hypergeometric
function $F(a_1,b_1,c_1;f)$, namely ${\bf Re}\,(c_1-a_1-b_1)>0$, which
clearly demands that we choose $\beta_1=\beta_1^{\,(-)}$.

We now turn to the radial regime close to the cosmological horizon $r_c$. 
Here, we will follow the previous studies \cite{Harmark, Crispino} and
approximate the metric function by 
%%%%%%%%%%%%%
\beq
h(r)=1-\biggl(\frac{r_h}{r}\biggr)^{(n+1)}
- \tilde\Lambda r^2\,\biggl[1-\biggl(\frac{r_h}{r}\biggr)^{(n+3)}\biggl]
\simeq 1- \tilde\Lambda r^2\,,
\label{ass_far}
\eeq
%%%%%%%%%
where we have again used Eq. (\ref{mu}) in order to eliminate the
$\mu$ parameter. Clearly, the validity of the above approximation
increases the farther away we move from the black-hole horizon, i.e.
the larger $r_c$ or the smaller $\Lambda$ is. In addition, the same
approximation becomes more accurate the larger the number of extra
dimensions is; therefore, we expect that this analytic approach
will lead to results with a more extended validity regime compared to
the ones derived in the purely 4-dimensional case.

If we then make the change of variable $r \rightarrow h(r) \simeq
1- \tilde \Lambda r^2$, the radial equation (\ref{radial_brane})
in the area close to $r=r_c$ takes the form
%%%%%%%%%%%
\begin{eqnarray}
h\,(1-h)\,\frac{d^2R}{dh^2} + \Bigl(1-\frac{5}{2}\,h\Bigr)\,\frac{dR}{dh} 
+ \frac{1}{4}\,\biggl[\,\frac{(\omega r_c)^2}{h}- \frac{l(l+1)}{(1-h)}
-\xi R^{(c)}_4 r_c^2\,\biggr] R=0\,,
\label{FF-1}
\end{eqnarray}
%%%%%%%%%%%%%%%
where now $R^{(c)}_4$ is the brane curvature (\ref{R4}) evaluated at
$r=r_c$. By making the field redefinition: 
$R(h)=h^{\alpha_2} (1-h)^{\beta_2} X(h)$, Eq. (\ref{FF-1}) takes again
the form of a hypergeometric equation (\ref{hyper}) where now the
various indices take the form
%%%%%%%%%
\beq
a_2=\alpha_2 + \beta_2 +\frac{3}{4} +\sqrt{\frac{9}{16}-\frac{\xi R^{(c)}_4 r_c^2}{4}}\,,
\eeq
%%%%%%%%%%%
\beq
b_2=\alpha_2 + \beta_2 + \frac{3}{4} -
\sqrt{\frac{9}{16}-\frac{\xi R^{(c)}_4 r_c^2}{4}}\,, \qquad c_2=1+ 2 \alpha_2\,.
\eeq
%%%%%%%%%%
The power coefficients $\alpha_2$ and $\beta_2$ are in this case determined
by the equations
%%%%%%%%%
\beq
\alpha_2^2 +\frac{\omega^2 r_c^2}{4} =0\,,
\label{alpha-dS}
\eeq
%%%%%%%%%
and
%%%%%%%%%
\beq
\beta_2^2 + \frac{\beta_2}{2} - \frac{l(l+1)}{4}=0\,, \label{beta-dS}
\eeq
%%%%%%%%%
respectively, and read 
%%%%%%%%%%%%%%
\begin{equation}
\alpha_2^{\,(\pm)} = \pm \frac{i \omega r_c}{2}\,, \qquad
\beta_2^{\,(\pm)} =\frac{1}{4}\,\Bigl[-1 \pm (2l+1)\Bigr]\,.
\label{alpha_beta}
\end{equation}
%%%%%%%%%%%%%

The general solution of the radial equation in the cosmological-horizon
regime can then be written as
%%%%%%%%
\begin{eqnarray}
&& \hspace*{-1cm}R_{C}(h)=B_1 \,h^{\alpha_2}\,(1-h)^{\beta_2}\,X(a_2,b_2,c_2;h)
\nonumber \\[1mm] && \hspace*{2cm} +\,
B_2\,h^{-\alpha_2}\,(1-h)^{\beta_2}\,X(a_2-c_2+1,b_2-c_2+1,2-c_2;h)\,,
\label{FF2-dS}
\end{eqnarray}
%%%%%%%%%
\noindent
where $B_{1,2}$ are again arbitrary constants. The criterion for the
convergence of the hypergeometric function $X(a_2,b_2,c_2;h)$, namely
${\bf Re}\,(c_2-a_2-b_2)>0$, forces us again to choose the negative sign
in the expression of the $\beta_2$ coefficient that then reads 
$\beta_2=-(l+1)/2$.

The metric function $h(r)$ goes again to zero when $r \rightarrow r_c$.
Then, the solution very close to the cosmological horizon reduces to
the expression
%%%%%%%%%%
\beq
R_{C}(h) \simeq B_1\,h^{\alpha_2} + B_2\,h^{-\alpha_2}\,. 
\end{equation}
%%%%%%
Since the effective potential vanishes also at $r_c$, the solution
is again expected to be comprised in that area by free plane waves.
Indeed, setting $h = e^{-2 r_*/r_c}$, the above asymptotic
solution may be re-written as
%%%%%%%%%%
\beq
R_{C}(r_*) \simeq B_1\,e^{\mp i\omega r_*} + B_2\,e^{\pm i\omega r_*}\,. 
\end{equation}
%%%%%%
Once again, the choice of the sign in the expression of $\alpha_2$, simply
interchanges the incoming and outgoing waves. In contrast to what happens
at the black-hole horizon, both waves are now allowed to exist at $r \simeq r_c$,
and it is their amplitudes that define the greybody factor for the emission of
scalar fields by the back hole. Thus, if we choose $\alpha_2=\alpha_2^{\,(+)}$, 
the greybody factor is simply given by
%%%%%%%%%%
\beq
|A|^2=1-\left|\frac{B_2}{B_1}\right|^2\,.
\label{greybody}
\end{equation}
%%%%%%%%%%%%

In order to complete the solution, we must ensure that the two asymptotic
solutions, $R_{BH}$ and $R_C$, can be smoothly matched at some arbitrary
intermediate value of the radial coordinate. Starting from the
near-black-hole solution, Eq. (\ref{BH-gen}) with $A_2=0$, we first shift
the argument of the hypergeometric function from $f$ to $1-f$, by using
the general relation \cite{Abramowitz}
%%%%%%%%%
\bea
\hspace*{-0.5cm}
F(a,b,c;x)&=&\frac{\Gamma(c)\,\Gamma(c-a-b)}
{\Gamma(c-a)\,\Gamma(c-b)}\,F(a,b,a+b-c+1;1-x) \nonumber \\[3mm]
&+& (1-f)^{c-a-b}\,\frac{\Gamma(c)\,\Gamma(a+b-c)}
{\Gamma(a)\,\Gamma(b)}\,F(c-a,c-b,c-a-b+1;1-x),
\label{stretching}
\eea
%%%%%%%%%
where $x$ is an arbitrary variable. Then, in the limit $r \gg r_h$,
or $f \rightarrow 1$, $R_{BH}$ takes the `stretched' form
%%%%%%%%%
\bea
R_{BH}(r) &\simeq& 
A_1\,\frac{\Gamma(c_1)\,\Gamma(a_1+b_1-c_1)}
{\Gamma(a_1)\,\Gamma(b_1)}\,\left(\frac{r}{r_h}\right)^{-(l+1)}+
A_1\,\frac{\Gamma(c_1)\,\Gamma(c_1-a_1-b_1)}
{\Gamma(c_1-a_1)\,\Gamma(c_1-b_1)}\,\left(\frac{r}{r_h}\right)^l
\nonumber \\[2mm]
&\equiv&  \Sigma_1\,r^{-(l+1)} +\Sigma_2\,r^l\,. \label{BH-stretched}
\eea
%%%%%%%%%
In the above, we have made the approximation that, for small
cosmological constant $\Lambda$ and coupling constant $\xi$, we may write
%%%%%%%%%%%
\bea
&&(1-f)^{\beta_1} \simeq \left(\frac{r_h}{r}\right)^{\beta_1 (n+1)}
\simeq \left(\frac{r}{r_h}\right)^l \\[1mm]
&&(1-f)^{\beta_1+c_1-a_1-b_1} \simeq \left(\frac{r_h}{r}\right)^{(2-B_h-\beta_1)(n+1)}
\simeq \left(\frac{r}{r_h}\right)^{-(l +1)}\,,
\eea
%%%%%%%%%%%%
since in that limit it holds that: $A_h \simeq (n+1)$, $B_h \simeq (2n+1)/(n+1)$
and $\beta_1 \simeq -l/(n+1)$. Note that the aforementioned approximations
are applied only in the expressions of the multiplying factors $(1-f)$
and not in the arguments of the Gamma functions in order to increase the
validity of our analytical results.

We now turn to the solution near the cosmological horizon (\ref{FF2-dS}).
Applying the same general relation (\ref{stretching}), in this case for
the variable $h$, we obtain
%%%%%%%%%
\bea
R_{C}(r) &\simeq& 
\left(\frac{r}{r_c}\right)^{-(l+1)}\,\left[B_1\,
\frac{\Gamma(c_2)\,\Gamma(c_2-a_2-b_2)}{\Gamma(c_2-a_2)\,\Gamma(c_2-b_2)}
+B_2\,\frac{\Gamma(2-c_2)\,\Gamma(c_2-a_2-b_2)}{\Gamma(1-a_2)\,\Gamma(1-b_2)}\right]
\nonumber \\[1mm]
&+& \left(\frac{r}{r_c}\right)^l \left[B_1\,\frac{\Gamma(c_2)
\,\Gamma(a_2+b_2-c_2)}{\Gamma(a_2)\,\Gamma(b_2)} +
B_2\,\frac{\Gamma(2-c_2)\,\Gamma(a_2+b_2-c_2)}
{\Gamma(a_2+1-c_2)\,\Gamma(b_2+1-c_2)}\right]
\nonumber \\[2mm]
&\equiv&  (\Sigma_3 B_1+\Sigma_4 B_2)\,r^{-(l+1)} +
(\Sigma_5 B_1+\Sigma_6 B_2)\,r^{l}\,. \label{C-stretched}
\eea
%%%%%%%%%
Here, we have made again the approximations 
%%%%%%%%%%%
\bea
&&(1-h)^{\beta_2} \simeq \left(\frac{r}{r_c}\right)^{2\beta_2}
= \left(\frac{r}{r_c}\right)^{-(l+1)} \\[1mm]
&&(1-h)^{\beta_2+c_2-a_2-b_2} \simeq \left(\frac{r}{r_c}\right)^{-(1+2\beta_2)}
= \left(\frac{r}{r_c}\right)^{l}\,,
\eea
%%%%%%%%%%%%
valid again for small cosmological constant $\Lambda$ and small coupling
constant $\xi$.

The two `stretched' solutions (\ref{BH-stretched}) and (\ref{C-stretched})
have the same power-law form and their smooth matching is straightforward:
identifying the coefficients of the same powers of $r$, we arrive at
the constraints
%%%%%%%%%%%%
\beq
B_1=\frac{\Sigma_1 \Sigma_6-\Sigma_2 \Sigma_4}
{\Sigma_3 \Sigma_6-\Sigma_4 \Sigma_5}\,, \qquad
B_2=\frac{\Sigma_2 \Sigma_3-\Sigma_1 \Sigma_5}
{\Sigma_3 \Sigma_6-\Sigma_4 \Sigma_5}\,.
\eeq
%%%%%%%%%%%%%
Then, it is easy to write the expression of the greybody factor
(\ref{greybody}) for the emission of scalar fields by a higher-dimensional
Scwarzschild-de-Sitter black hole on the brane; this has the form
%%%%%%%%%%%%%
\beq
|A^2|=1-\left|\frac{\Sigma_2 \Sigma_3-\Sigma_1 \Sigma_5}
{\Sigma_1 \Sigma_6-\Sigma_2 \Sigma_4}\right|^2\,.
\label{grey-brane-final}
\eeq
%%%%%%%%%%%%%%

A few comments are in order at this point regarding the approximations
made in our analysis. The assumed range of values of the variables $f$
and $h$ as well as their approximated forms are indeed highly accurate
as long as the value of the cosmological constant is small and the
distance between $r_h$ and $r_c$ is large. As soon as $\Lambda$ becomes
large, deviations from the expected behaviour for the greybody factor
will start to appear. We will explicitly demonstrate this in section
3.3 and thus define the range of validity of our analytical results.
Finally, let us stress that the aforementioned approximations, necessary
for the smooth matching of the two asymptotic solutions, do not involve
at all the energy of the emitted particle. To our knowledge, this is
in contrast with all previous similar studies where a low-energy condition
was always imposed. This creates the expectation that our
analytic result may be valid beyond the low-energy regime.
We will return again to this point in section 3.3.

%%%%%%%%%%%%%%%%%%%%%%%%%%%%%%%%%%%%%%%%%

\subsection{The Low-energy Limit}

In this subsection, we will derive a simplified expression for the greybody
factor (\ref{grey-brane-final}) in the low-energy limit, i.e.
when $\omega \rightarrow 0$, in the case of both minimal and non-minimal coupling. 
In the former case, we will focus on the dominant partial wave with $l=0$,
and demonstrate that our expression correctly reproduces the non-vanishing
asymptotic value derived previously in the literature \cite{KGB}. 
In the latter case, we will analytically show that, in the limit
$\omega \rightarrow 0$, the aforementioned asymptotic value disappears
as does also the ${\cal O}(\omega)$ term in the expansion of the
greybody factor. The first non-vanishing term of ${\cal O}(\omega^2)$
for arbitrary $l$ and coupling parameter $\xi$ will be presented.

We will present a unified analysis for scalars with minimal and non-minimal
coupling by keeping the parameter $\xi$ arbitrary, and consider
particular values at a later stage. For the low-energy expansion, it is
convenient to re-write the ($a_i,b_i,c_i)$ parameters of the hypergeometric functions
in a way that separates the $\omega$-dependence. To this end, we define the
$\omega$-independent quantities
%%%%%%%%%%%%%%%
\beq
\delta= \frac{1+ 2n - \sqrt{1+4 \lambda_h}}{2 (n+1)}\,,
\label{gamma}
\eeq
%%%%%%%%%%%%
\beq
\epsilon= \frac{1 - \sqrt{1+4 \lambda_h}}{2 (n+1)}\,,
\label{delta}
\eeq
%%%%%%%
\beq
\eta_{\pm}= \frac{1}{4}\left(1-2l \pm \sqrt{9-4 \xi R_4^{(c)}r_c^2}\right)\,,
\label{eta}
\eeq
%%%%%%%%
and write
%%%%%%%%%%%%%%%
\beq
a_1=\delta -\frac{i \omega r_h}{(n+1)}\,,
\quad b_1=\epsilon-\frac{i \omega r_h}{(n+1)}\,, \quad
c_1=1-\frac{2 i \omega r_h}{(n+1)}\,, \label{abc_rh}
\eeq
%%%%%%%%%%%%%%%%%
for the parameters near the black-hole horizon, and
%%%%%%%%%%%%%%%
\beq
a_2=\eta_+ +\frac{i \omega r_c}{2}\,, \quad 
b_2=\eta_- +\frac{i \omega r_c}{2}\,, \quad
c_2=1+i \omega r_c\,, \label{abc_rc}
\eeq
%%%%%%%%%%%%%
for the ones close to the cosmological horizon. Note that above, in order
to simplify the fairly complex calculation, we have also taken the limit
of small cosmological constant apart from the terms where $\xi$ is involved.

Then, the $\Sigma_i$ quantities, defined in Eqs. (\ref{BH-stretched}) and
(\ref{C-stretched}), can be written in the form
%%%%%%%%%%%%%%%%%
\beq
\Sigma_1=\frac{r_h^{l+1}\,\Gamma(1-2i \omega R_H)\,\Gamma(\delta+\epsilon-1)}
{\Gamma(\delta-i\omega R_H)\,\Gamma(\epsilon-i \omega R_H)}\,,
\eeq
%%%%%%%%%%%%%%
\beq
\Sigma_2=\frac{ r_h^{-l}\,\Gamma(1-2 i \omega R_H)\,\Gamma(1-\delta-\epsilon)}
{\Gamma(1-\delta -i \omega R_H)\,\Gamma(1-\epsilon - i \omega R_H)}\,,
\eeq
%%%%%%%%%%%%%%%
\beq
\Sigma_3=\frac{r_c^{(l+1)}\,\Gamma(1+2 i \omega R_C)\,\Gamma(1-\eta_+-\eta_-)}
{\Gamma(1-\eta_+ +i \omega R_C)\,\Gamma(1-\eta_- +i \omega R_C)}\,,
\eeq
%%%%%%%%%%%%%
\beq
\Sigma_4=\frac{r_c^{(l+1)}\,\Gamma(1-2 i \omega R_C)\,\Gamma(1-\eta_+ -\eta_-)}
{\Gamma(1-\eta_+ -i \omega R_C)\,\Gamma(1-\eta_- -i \omega R_C)}=\overline{\Sigma_3}\,,
\eeq
%%%%%%%%%%%
\beq
\Sigma_5=\frac{r_c^{-l}\,\Gamma(1+2 i \omega R_C)\,\Gamma(\eta_+ +\eta_- -1)}
{\Gamma(\eta_+ +i \omega R_C)\,\Gamma(\eta_- +i \omega R_C)}\,,
\eeq
%%%%%%%%%%%
\beq
\Sigma_6=\frac{r_c^{-l}\,\Gamma(1-2 i \omega R_C)\,\Gamma(\eta_+ +\eta_- -1)}
{\Gamma(\eta_+ -i \omega R_C)\,\Gamma(\eta_- -i \omega R_C)} = \overline{\Sigma_5}\,,
\eeq
%%%%%%%%%%%
where we have also defined $R_H \equiv r_h/(n+1)$ and $R_C \equiv r_c/2$. 

We first consider the minimal coupling case, and focus on the lowest, dominant
mode of the scalar field. Upon setting $\xi=0$ and $l=0$, the quantities in
Eqs. (\ref{gamma})-(\ref{eta}) are greatly simplified and read
%%%%%%%%%%%%
\beq
\delta=\frac{n}{n+1}\,, \qquad \epsilon=0\,, \qquad
\eta_\pm=\left(1,-\frac{1}{2}\right)\,. 
\eeq
%%%%%%%%%%%%%%%
Then, the $\Sigma_i$ quantities also assume simpler forms that may easily be
expanded in power series in the energy $\omega$. For example, $\Sigma_1$
is written as
%%%%%%%%%%%%
\beq
\Sigma_1 \approx \frac{r_h\,\Gamma(1-\frac{2 i \omega r_h}{n+1})\,\Gamma(-\frac{1}{n+1})}
{\Gamma(\frac{n-i \omega r_h}{n+1})\,\Gamma(-\frac{i \omega r_h}{n+1})}
\rightarrow i \omega r_h^2 + {\cal O} (\omega^2)\,.
\eeq
%%%%%%%%%%%%
In the above, we have applied the Gamma function property: 
$z\,\Gamma[-z]-=-\Gamma [1-z]$, and used the expansion formulae
\cite{Abramowitz}
%%%%%%%%%%%%%
\bea
\Gamma(\hat a + i \omega \hat b)&=&\Gamma(\hat a)\,\left[1+i\omega \hat b
\Psi ^{(0)}(\hat a)\right]
+{\cal O}(\omega^2)\,, \\[1mm]
\Gamma(i \omega \hat b) &=&\frac{1}{i \omega \hat b} -\gamma +{\cal O}(\omega)\,,
\eea
%%%%%%%%%%%%
where $\hat a$ and $\hat b$ are arbitrary $\omega$-independent quantities,
$\gamma$ is the Euler's constant, and $\Psi ^{(0)}$ the poly-gamma function. 
Similar expressions follow for the other $\Sigma_i$ quantities, namely
%%%%%%%%%%%%%%
\beq
\Sigma_2 \approx 1+\frac{i \omega r_h}{n+1} \left[\gamma + 
\Psi^{(0)}\left(\frac{1}{n+1}\right) \right]+ {\cal O} (\omega^2)\,,
\eeq
%%%%%%%%%%%%
\beq
\Sigma_3 \approx i \omega r_c^2 + {\cal O} (\omega^2)=\overline{\Sigma_4}\,,
\eeq
%%%%%%%%%%%%%%
\beq
\Sigma_5 \approx 1+i \omega r_c\,(\log{2}-1)+ {\cal O}(\omega^2)
=\overline{\Sigma_6}\,.
\eeq
%%%%%%%%%%%%%%%
Under these approximations we finally obtain, from Eq. (\ref{grey-brane-final}),
the result
%%%%%%%%%%%%%%%%%%%%
\beq
|A^2|=1-\left|\frac{i \omega\,(r_c^2-r_h^2)}{i \omega\,(r_c^2+r_h^2)}\right|^2
+{\cal O}(\omega) =
\frac{4 r_h^2r_c^2}{(r_c^2+r_h^2)^2}+ {\cal O}(\omega)\,.
\label{geom_brane}
\eeq
%%%%%%%%%%%%%%
Therefore, our general expression (\ref{grey-brane-final}) correctly reproduces
the low-energy geometric value of the greybody factor for the mode $l=0$, in
accordance to previous analyses \cite{KGB,Harmark,Crispino}. This feature is
characteristic of the propagation of free, massless scalar particles in a 
Schwarzschild-de-Sitter spacetime, both four- and higher-dimensional, and
leads to the enhanced emission of very soft modes in the Hawking radiation
spectra \cite{KGB}. 

Let us now assume that the non-minimal coupling parameter $\xi$ takes an arbitrary
non-vanishing value. In this case, we observe that all $\Sigma_i$ quantities
are proportional to the following combination:
\begin{equation}
\frac{\Gamma(1 \pm 2 i \omega R)}{\Gamma(\hat a \pm i \omega R)\,\Gamma(\hat b \pm i \omega R)}
= \frac{1 \mp i \omega R \left[2 \gamma + \Psi^{(0)}(\hat a)+\Psi^{(0)}(\hat b)  \right]}
{\Gamma(\hat a)\,\Gamma(\hat b)}+ {\cal O}(\omega^2)\,, 
\label{combination}
\end{equation}
%%%%%%%%%%%%%%%%
where $R$ is either $R_H$ or $R_C$, $\hat a$ and $\hat b$ are again arbitrary, but
non-vanishing, $\omega$-independent quantities. Then, using the above expansion
formula, we find that
%%%%%%%%%%%%%%%%
\bea
\Sigma_1 \Sigma_5 &=& K \left(1-i \omega R_C B+i\omega R_H \Gamma\right),
\nonumber \\[1mm]
\Sigma_1 \Sigma_6 &=& K \left(1+i \omega R_C B +i\omega R_H \Gamma\right),\\[1mm]
\Sigma_2 \Sigma_3 &= & E \left(1+i \omega R_H Z -i \omega R_C \Theta \right),
\nonumber\\[1mm]
\Sigma_2 \Sigma_4 &=& E \left(1+i \omega R_H Z+i \omega R_C \Theta \right),
\nonumber
\eea
%%%%%%%%%%%%%
where
%%%%%%%%%%%
\bea
K &\equiv& \frac{r_h^{(l+1)} r_c^{-l}\,
\Gamma(\delta+\epsilon-1)\,\Gamma(\eta_+ +\eta_- -1)}
{\Gamma(\delta)\,\Gamma(\epsilon)\,\Gamma(\eta_+)\,\Gamma(\eta_-)}\,,
\nonumber \\[1mm]
E &\equiv& \frac{r_h^{-l} r_c^{l+1}\,
\Gamma(1-\delta -\epsilon)\,\Gamma(1-\eta_+ -\eta_-)}
{\Gamma(1-\delta)\,\Gamma(1-\epsilon)\,\Gamma(1-\eta_+)\Gamma(1-\eta_-)}\,,
\eea
%%%%%%%%%%%%
and
%%%%%%%%%%%%%
\bea
B &\equiv& 2 \gamma+ \Psi^{(0)}(\eta_+)+\Psi^{(0)}(\eta_-)\,,\nonumber \\[2mm]
%%%
\Gamma &\equiv& 2\gamma+\Psi^{(0)}(\delta)+\Psi^{(0)}(\epsilon)\nonumber \\[2mm]
Z &\equiv& 2 \gamma+ \Psi^{(0)}(1-\delta)+ \Psi^{(0)}(1-\epsilon)\,,\\[2mm]
\Theta &\equiv& 2 \gamma+ \Psi^{(0)}(1-\eta_+)+ \Psi^{(0)}(1-\eta_-)\,. \nonumber
\eea
%%%%%%%%%%%%%%
Then, we easily find
%%%%%%%%%%%%
\bea
\Sigma_2 \Sigma_3- \Sigma_1 \Sigma_5 &=&
(E-K)+i \omega R_H\,(E Z -K \Gamma)+i \omega R_C\,(K B -E \Theta)\,,\nonumber\\[2mm]
\Sigma_1 \Sigma_6 -\Sigma_2 \Sigma_4 &=&
(K-E)+i \omega R_H\,(K \Gamma-E Z)+i \omega R_C\,(K B -E \Theta)\,,
\eea
%%%%%%%%%%%%%%
from which it follows that
%%%%%%%%%%%%%%
\beq
|\Sigma_2 \Sigma_3- \Sigma_1 \Sigma_5|^2 \simeq
|\Sigma_1 \Sigma_6 -\Sigma_2 \Sigma_4|^2= (K-E)^2 +{\cal O}(\omega^2)\,.
\eeq
%%%%%%%%%%%%%
Substituting the above result in Eq. (\ref{grey-brane-final}), we find that
the first non-vanishing term in the low-energy expansion of the greybody
factor, in the case where $\xi \neq 0$, is the one of ${\cal O}(\omega^2)$.
This holds for all partial waves including the dominant one with $l=0$.
Therefore, in the presence of a non-minimal coupling to gravity, there are
no scalar modes with a non-vanishing low-energy asymptotic value of their
greybody factor.

The exact expression of the ${\cal O}(\omega^2)$ term for arbitrary $l$ is
extremely difficult to derive analytically. Therefore, we merely display
here its expression derived via symbolic calculation; it has the form
%%%%%%%%%%%%%%
\bea
|A|^2=  \frac{8 \pi^2 (\omega r_h)^2 \lambda_h^l \left[\Gamma(\theta_{+})
\,\Gamma(\theta_{-})\right]^2 \Gamma[\frac{1+u}{2 (n+1)}]\,
\Gamma[\frac{1+2n+u}{2 (n+1)}]}{(1+2l)u \left( \cos[\frac{n \pi}{(n+1)}]+
\cos[\frac{\pi u}{(n+1)}] \right)\Gamma[\frac{1}{2}+l]^2\,\Gamma[\frac{u}{(n+1)}]^2
\,\Gamma[\frac{1+2n-u}{2(n+1)}]\,\Gamma[\frac{1-u}{2 (n+1)}]}
\label{Crispino's generalization}
\eea
%%%%%%%%%%%%%%%
where $u \equiv \sqrt{(2l+1)^2+ 4 \xi R_4^{(h)}  r_h^2}$, and 
%%%%%%%%%%%%%
\beq
\theta_{\pm}=\frac{1}{4} \left( 3+2 l \pm \sqrt{9-4 \xi R_4^{(c)}r_c^2}\right)\,.
\label{theta}
\eeq
%%%%%%%%%%%
As an additional consistency check, we have confirmed that the above expression
reduces\footnote{The agreement is almost perfect since our result differs by
the one appearing in \cite{Crispino} by only a factor $4^l$ caused perhaps by
a typographical error.}, after setting $n=0$, to the one derived in
\cite{Crispino} in the context of the four-dimensional analysis.

%%%%%%%%%%%%%%%%%%%%%%%%%%%%%%%%%%%%%%%%%%%%%%%%%%%%%%%%%%%%%%

\subsection{Plotting our analytic result}

We are now ready to investigate in detail the complete form of the greybody
factor for the emission of scalar fields on the brane by a higher-dimensional
Schwarzschild-de-Sitter black hole, as this is given by the analytic
expression (\ref{grey-brane-final}). We will also study its dependence
on both particle ($l$, $\xi$) and spacetime properties ($n$, $\Lambda$).

\begin{figure}[t]
  \begin{center}
\mbox{\includegraphics[width = 0.42 \textwidth] {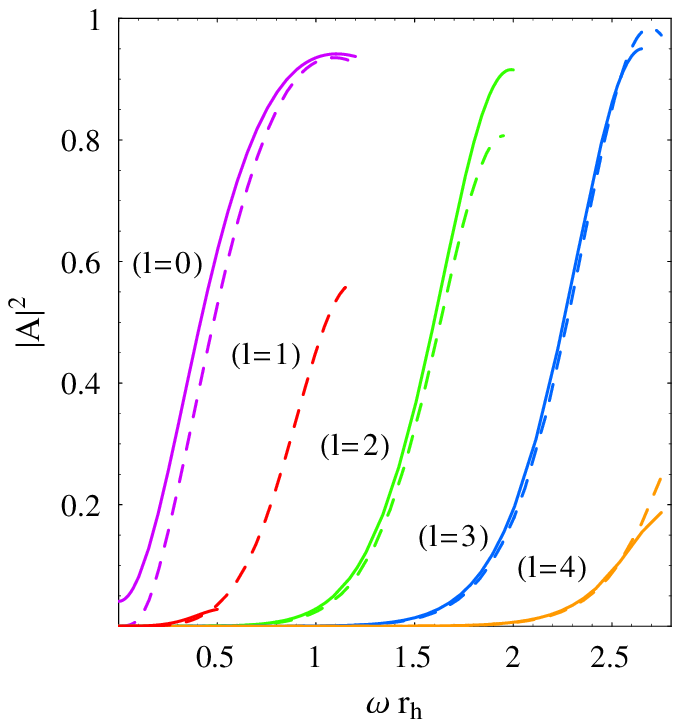}}
\hspace*{0.5cm} {\includegraphics[width = 0.42 \textwidth] {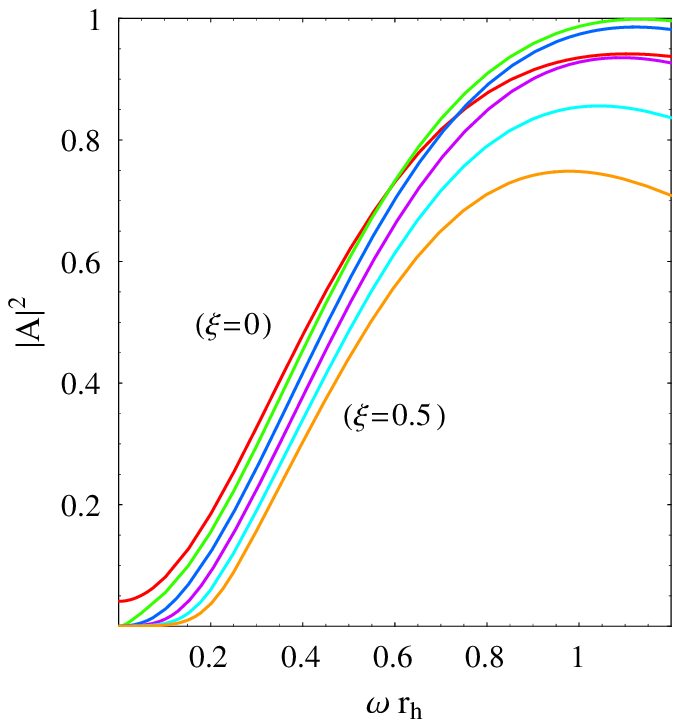}}
    \caption{ Greybody factors for brane scalar fields, for $n=2$ and
	$\Lambda=0.1$, and: \textbf{(a)} for variable $l=0,1,2,3,4$ and $\xi=0$
	(solid lines) or $\xi=0.3$ (dashed lines); \textbf{(b)} for $l=0$ and
	variable $\xi=0,0.1,0.2,0.3,0.4,0.5$.}
   \label{grey_brane_ellxi}
  \end{center}
\end{figure}

In Fig. \ref{grey_brane_ellxi}, we depict the greybody factor $|A|^2$ as a
function of the dimensionless energy parameter $\omega r_h$. The graph on
the left, Fig. \ref{grey_brane_ellxi}(a), shows the behaviour of $|A|^2$
for the first five partial waves with $l=0,1,2,3,4$, for $n=2$ and
$\Lambda=0.1$ (in units of $r_h^{-2}$); for every value of the
angular-momentum number $l$, the solid line corresponds to the case of
minimal coupling, with $\xi=0$, and the dashed line to an arbitrary
non-minimal coupling, with $\xi=0.3$. As expected from the dependence
of the effective potential on $l$, the greybody factor gets suppressed
for large $l$, thus rendering the lowest partial wave with $l=0$ the
dominant one; this holds independently of whether the scalar field is
coupled minimally or non-minimally to gravity. The presence of $\xi$
causes a significant modification to the behaviour of $|A|^2$ only for
the $l=0$ partial wave: in accordance to the discussion in the previous
subsection, for $\xi=0$ we recover the non-vanishing low-energy 
geometric value of the greybody factor (\ref{geom_brane}), whereas
for $\xi \neq 0$ this limit becomes zero.

%%%%%%%%%%%%%%%%%%%%%
%%%%%%%%%%%%%%%%%  

We note that, in order to be able to display the behaviour of all five
partial waves in the same plot, we were forced to extend the energy range
well beyond the low-energy regime, where $\omega r_h \ll 1$. For most
partial waves shown, almost the whole of the curve of the greybody factor,
extending in principle from zero to unity, is also visible. 
This is rather unusual
in results following from an analytical approach, and, as we commented
at the end of Section 3.1, this may be due to the fact that the
energy parameter was never involved in the approximations made. 
As the low-energy assumption is always the reason that the analytic
results deviate from the exact numerical ones as the energy increases,
we envisage that in the present case our results may be fairly close
to the exact ones even for large values of energy. We should however
add that our analytic results still show occasional signs of defects
caused by the existence of poles in the expression of Gamma functions
-- this is an inherent feature of the analytic approach used in general
in studying Hawking radiation in black-hole spacetimes which employs
hypergeometric functions. It is the existence of such a pole that
causes the abrupt stop of the solid line for the partial wave with
$l=1$ in Fig. \ref{grey_brane_ellxi}(a) at a fairly small value of energy.
Although the presence of the non-minimal coupling parameter $\xi$
apparently affects only marginally the behaviour of $|A|^2$ for
the case $l \neq 0$, its presence shifts the value of the energy 
where the pole emerges: it is for this reason that the dashed curve
for the partial wave with $l=1$ manages to extend up to the intermediate
energy regime. 

In Fig. \ref{grey_brane_ellxi}(b), we display the dependence of the
greybody factor on the other particle parameter, its coupling parameter
$\xi$: for the lowest partial wave with $l = 0$, $n = 2$ and $\Lambda = 0.1$,
$|A|^2$ is plotted for $\xi = 0,0.1,0.2,0.3,0.4,0.5$. The transition from
the non-zero low-energy asymptotic value to a zero one, as $\xi$ assumes
a non-vanishing value, is again evident. We also observe that the greybody
factor is suppressed as the value of the coupling parameter increases,
again in accordance with the behaviour of the effective potential.
This suppression is similar to the one caused by the presence of a
mass term for the scalar field that has been found before in the
literature in different black-hole spacetimes 
\cite{Page, Jung, Sampaio, KP1} -- indeed, the way the non-minimal
coupling term appears in the scalar equation of motion (\ref{field-eq-brane})
resembles the one of a mass term. Also, in the purely four-dimensional
analysis of \cite{Crispino}, it was 
proposed that this resemblance between the two terms is the one that
causes the vanishing of the low-energy asymptotic value of the greybody
factor when $\xi \neq 0$: the geometric asymptotic value (\ref{geom_brane})
is characteristic of a free, massless particle, as shown in \cite{KGB},
while a massive scalar particle does not exhibit such a feature. 
The same behaviour noted in \cite{Crispino} is also observed in the
present higher-dimensional set-up, a result that adds further support
to this argument.

%%%%%%%%%%%%%%%%%%%%%
\begin{figure}[t]
  \begin{center}
\mbox{\includegraphics[width = 0.42 \textwidth] {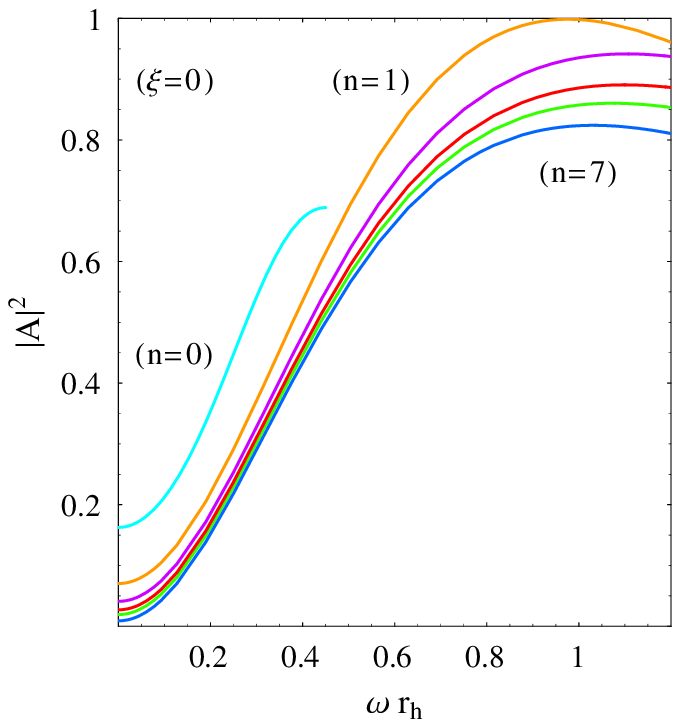}}
\hspace*{0.5cm} {\includegraphics[width = 0.42 \textwidth] {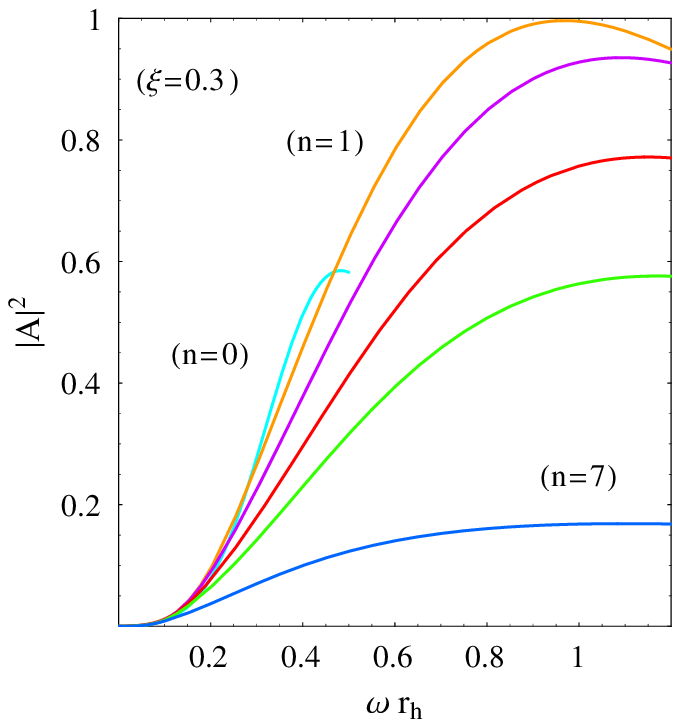}}
    \caption{ Greybody factors for brane scalar fields, for $l=0$, 
	$\Lambda=0.1$, and variable $n=0,1,2,3,4,7$ and: \textbf{(a)} for $\xi=0$,
	and \textbf{(b)} $\xi=0.3$.}
   \label{grey_brane_n}
  \end{center}
\end{figure}
%%%%%%%%%%%%%%%%%

Next, we turn to the dependence of the greybody factor on the spacetime
parameters. In Fig. \ref{grey_brane_n}(a,b), we show the behaviour of 
$|A|^2$ in terms of the number of extra dimensions $n$, for minimal
($\xi=0$) and non-minimal ($\xi=0.3$) coupling, respectively -- the
remaining parameters have been set to $l=0$ and $\Lambda=0.1$. Although
the scalar field is not free to propagate in the bulk, the projected
spacetime on the brane (\ref{metric_brane}) depends on $n$, and this is
reflected in the profile of the greybody factor. For $\xi =0$, and in
the low-energy limit, $|A|^2$ goes to its geometric value (\ref{geom_brane}),
as expected. Although this expression does not have either an explicit
dependence on $n$, it has an implicit one since the location of $r_h$
and $r_c$, the roots of the metric function $h(r)$ (\ref{h-fun}), depends
on $n$. Fig. \ref{grey_brane_n}(a) shows that, the asymptotic geometric value
decreases, as $n$ increases from 0 to 7; this is due to the fact that,
for fixed  $\Lambda$, the increase in the value of $n$ causes the two horizons
to move farther apart -- this increases the size of the causal spacetime
that the scalar field needs to traverse and decreases the corresponding
greybody factor. The decrease with $n$ characterizes the behaviour of
$|A|^2$ over the whole energy regime, and the same effect persists also
for non-zero $\xi$, as one may see in Fig. \ref{grey_brane_n}(b). In
fact, the suppression here is much more prominent, especially for large
values of $n$.

%%%%%%%%%%%%%%%%%%%
\begin{figure}[t]
  \begin{center} \hspace*{-0.3cm}
  \begin{tabular}{ccc}
\includegraphics[width = 0.33 \textwidth] {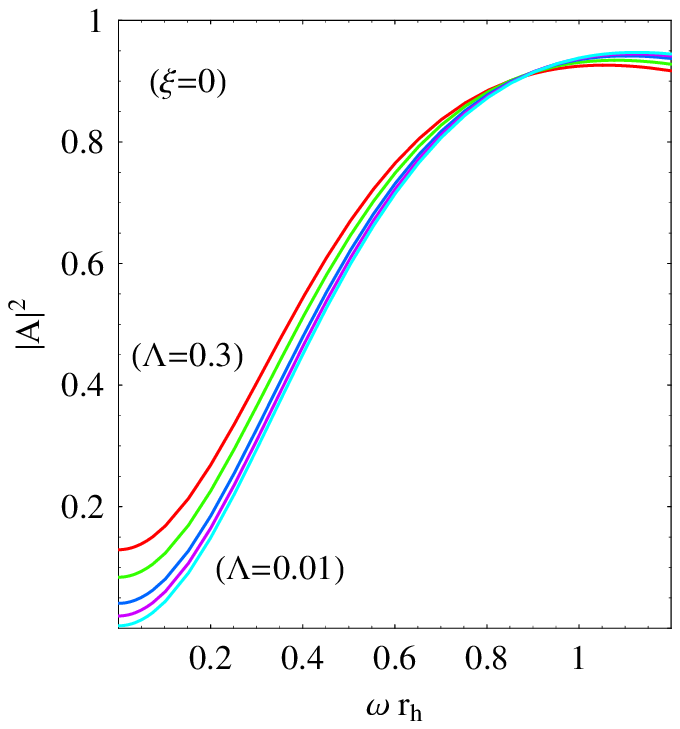}
&
\hspace*{-0.5cm} {\includegraphics[width = 0.33 \textwidth]
{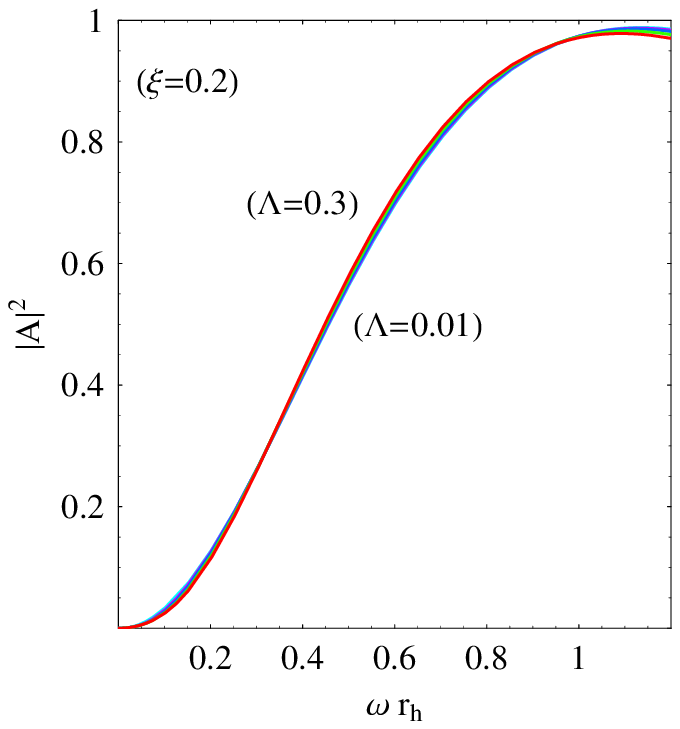}}&
\hspace*{-0.5cm} {\includegraphics[width = 0.33 \textwidth]
{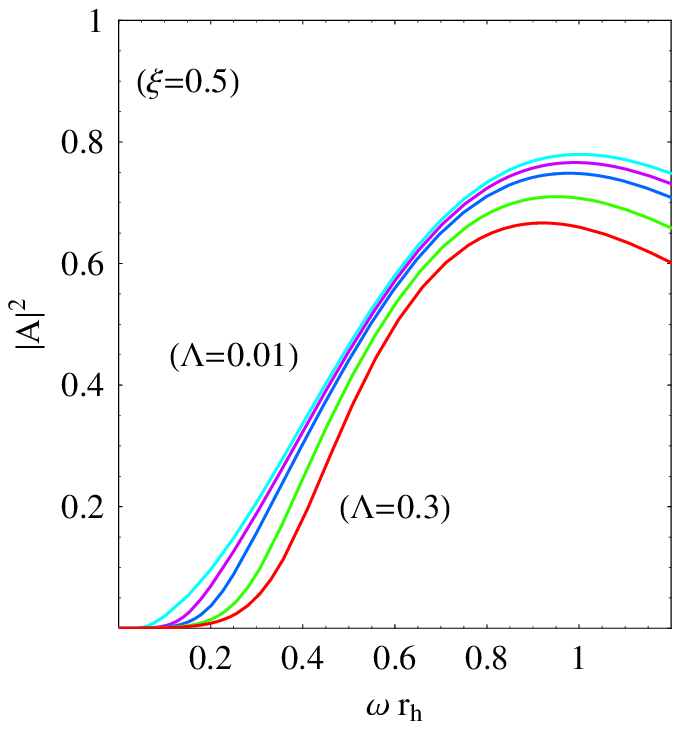}} \end{tabular}
    \caption{ Greybody factors for brane scalar fields for $l=0$, $n=2$, and
	$\Lambda=$0.01, 0.05, 0.1, 0.2, 0.3, and \textbf{(a)} for $\xi=0$,
	\textbf{(b)} $\xi=0.2$, and \textbf{(c)} $\xi=0.5$.}
   \label{grey_brane_Lam}
  \end{center}
\end{figure}
%%%%%%%%%%%%%%%%%

Finally, in Fig. \ref{grey_brane_Lam}, we depict the dependence of the
greybody factor on the bulk cosmological constant, for three different
values of the non-minimal coupling parameter, $\xi=0,0.2$ and 0.5, for
the dominant partial wave with $l=0$ and for $n=2$. For $\xi=0$, the value
of $|A|^2$ is clearly enhanced as $\Lambda$ gradually takes on the values 0.01,
0.05, 0.1, 0.2 and 0.3 - the enhancement seems to disappear beyond the
intermediate energy regime but the validity of our analytic approach
may also be in question there. As $\xi$ increases, the value of the
greybody factor becomes less sensitive to the value of $\Lambda$, and
eventually, for $\xi=0.2$ (as shown in Fig. \ref{grey_brane_Lam}(b)),
$|A|^2$ is almost independent of it. For even higher values of the
coupling parameter, the situation is reversed: now, the greybody
factor is suppressed as $\Lambda$ increases (see Fig. \ref{grey_brane_Lam}(c)).
This behaviour is the
result of the competition between two different $\Lambda$-contributions:
one in the metric function, that works towards subsidizing the energy
of the emitted particle thus giving a boost to its greybody factor
\cite{KGB}, and one in the non-minimal coupling term that, acting as a
mass term, suppresses its emission probability. For zero or very small
values of $\xi$, the latter effect is negligible and the former 
dominates; for intermediate values, the two effects cancel each other,
while for large values of $\xi$, the latter effect is the most important. 

As shown in Fig. \ref{grey_brane_Lam}(a), for the case of minimal coupling,
i.e. for $\xi=0$, the greybody factor of the dominant $l=0$ partial wave 
adopts a low-energy non-zero asymptotic value. Provided that the
approximations considered in our analysis are respected, this asymptotic
value should be given by the one in Eq. (\ref{geom_brane}) - this value has
been confirmed by exact numerical analyses both in a higher-dimensional
\cite{KGB} and four-dimensional context \cite{Crispino}. However, for
large values of the cosmological constant, i.e. larger than $\Lambda=0.3$,
the low-energy asymptotic value of $|A|^2$, as given by the expression
(\ref{grey-brane-final}), deviates from the exact one (\ref{geom_brane})
by more than 10\% while for $\Lambda \leq 0.1$ the error falls below 4\%.
This is due to the fact that, as $\Lambda$ increases, some of our
approximations, such as (\ref{ass_far}), are not satisfied anymore. 
Demanding that the deviation between the derived and expected asymptotic
limits is smaller than 5\% sets the allowed range of values of the
cosmological constant that may be considered to $\Lambda \leq 0.1$.
It is worth noting that, even when $\Lambda$ takes a large value,
increasing the value of $n$ restores the accuracy of the analysis;
this is, for example, obvious from Eq. (\ref{ass_far}) where the terms
that need to be ignored become indeed more negligible the larger $n$ is.

%%%%%%%%%%%%%%%%%%%%%%%%%%%%%%%%%%%%%%%%%%%%%%%%%%%%%%%%%%%%%%%%%%%%%%%%%%%
%
%%%%%%%%%%%%%%%%%%%%%%%%%%%%%%%%%%%%%%%%%%%%%%%%%%%%%%%%%%%%%%%%%%%%%%%%%%%

\section{Emission of Scalar Particles in the Bulk} 

We will now consider the case of the emission of scalar particles in
the bulk that may be again minimally or non-minimally coupled to
gravity. Then, the higher-dimensional action (\ref{action_D}) is
supplemented by the scalar part 
%%%%%%%%%%%%%%%
\beq
S_\Phi=-\frac{1}{2}\,\int d^{4+n}x \,\sqrt{-G}\left[\xi \Phi^2 R_D
+\partial_M \Phi\,\partial^M \Phi \right]\,.
\label{action-scalar-bulk}
\eeq
%%%%%%%%%%%%%%
where now $G_{MN}$ is the higher-dimensional metric tensor defined in
Eq. (\ref{bhmetric}), and $R_D$ the corresponding curvature given in
Eq. (\ref{RD}). 

The equation of motion of the bulk scalar field now reads
%%%%%%%%%%%%
\beq
\frac{1}{\sqrt{-G}}\,\partial_M\left(\sqrt{-G}\,G^{MN}\partial_N \Phi\right)
=\xi R_D\,\Phi\,. 
\label{field-eq-bulk}
\eeq
%%%%%%%%%%%%%
We assume the factorized ansatz
%%%%%%%%
\beq \Phi(t,r,\theta_i,\varphi) = e^{-i\omega t}R(r)\,\tilde Y(\theta_i,\varphi)\,,
\eeq
%%%%%%%%
where $\tilde Y(\theta_i,\varphi)$ are the hyperspherical harmonics
\cite{Muller}. Their equation is the $(n+2)$-dimensional generalisation
of the usual two-dimensional one for the scalar harmonics, with eigenvalue
$l(l+n+1)$. The angular and radial part are again decoupled with the equation
for the radial function $R(r)$ given by
%%%%%%%%%%
\beq
\frac{1}{r^{n+2}}\,\frac{d \,}{dr} \biggl(hr^{n+2}\,\frac{d R}{dr}\,\biggr) +
\biggl[\frac{\omega^2}{h} -\frac{l(l+n+1)}{r^2}-\xi R_D\biggr] R=0\,.
\label{radial_bulk}
\eeq
%%%%%%%%%%%

\begin{figure}[t]
  \begin{center}
\mbox{\includegraphics[width = 0.5 \textwidth] {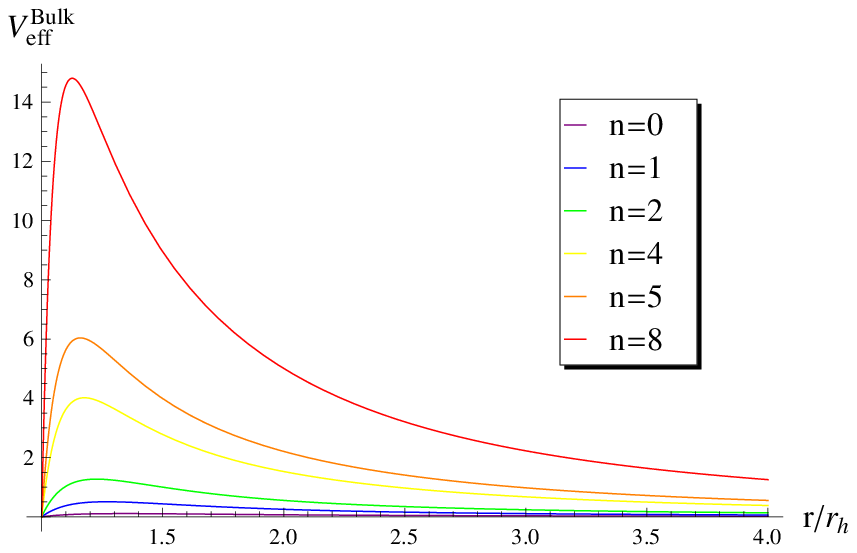}}
\hspace*{-0.4cm} {\includegraphics[width = 0.5 \textwidth]
{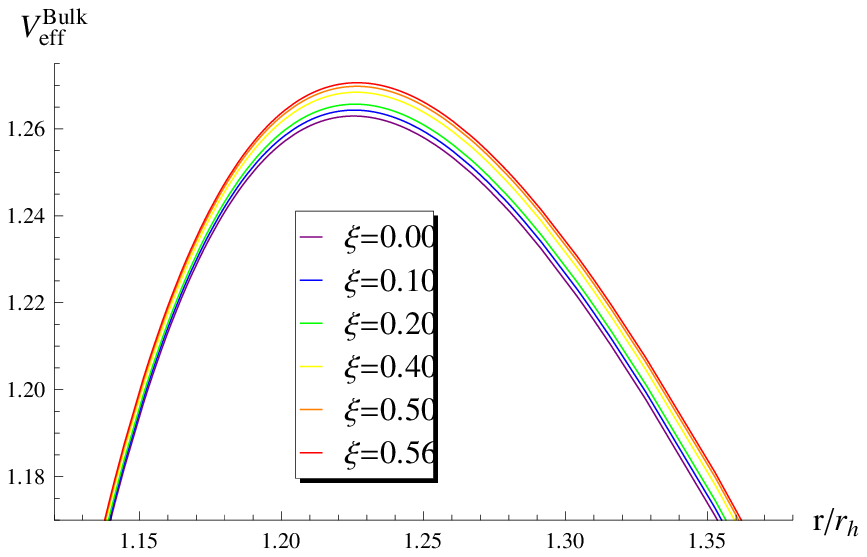}}
    \caption{Effective potential for bulk scalar fields for: \textbf{(a)}
	$l=0$, $\Lambda=0.01$, $\xi=0.5$ and variable $n=0,1,2,4,5,8$ (from bottom
	to the top), and \textbf{(b)} $l=0$, $\Lambda=0.01$, $n=2$ and variable
	$\xi=0,0.1,0.2,0.4,0.5,0.56$ (again, from bottom to top).}
   \label{pot_bulk_nxi}
  \end{center}
\end{figure}

If we redefine the radial function as $u(r)=r^{(n+2)/2} R(r)$ and use
again the tortoise coordinate $dr_*=dr/h(r)$, then, Eq. (\ref{radial_bulk})
may be re-written in the Schr\"odinger-like form where now the effective
potential reads
%%%%%%%%%%%%
\beq
V_{\rm eff}^{\rm bulk}=h(r) \left[\frac{l(l+n+1)}{r^2}+\xi R_D +
\frac{(n+2)}{2r}\,\frac{dh}{dr}+\frac{n(n+2)h}{4r^2}\right]\,,
\eeq
%%%%%%%%%%%%
or more explicitly
%%%%%%%%%%%%
\beq
V_{\rm eff}^{\rm bulk}=h(r) \left\{\frac{(2l+n+1)^2-1}{4r^2}+
\kappa^2_D \Lambda\,(n+4) \left[\frac{2\xi}{(n+2)}-\frac{1}{2(n+3)}\right]
 + \frac{(n+2)^2 \mu}{4r^{n+3}}\right\}\,.
 \label{pot_bulk}
\eeq
%%%%%%%%%%%%
The parameter $\mu$ can again be eliminated by using Eq. (\ref{mu}) and the
black-hole horizon fixed at an arbitrary value. As is obvious from the above,
the bulk effective potential vanishes at the two horizons, $r_h$ and $r_c$,
and its height increases again with the angular-momentum number $l$. Its
profile in terms of the number of extra dimensions $n$ and coupling
parameter $\xi$ is shown in Figs. \ref{pot_bulk_nxi}(a,b), respectively.
As in the case of propagation on the brane, the gravitational barrier in
the bulk also rises as either $n$ of $\xi$ increases. We note that the 
height of the barrier in the bulk is higher than the one on the brane
which points to the dominance of brane emission over the bulk one, as
noted also before for the case of minimal coupling \cite{KGB}. Also,
the bulk potential is much less sensitive to the value of the non-minimal
coupling parameter, and this is why a magnification of the area around
the peak of the potential is shown in Fig. \ref{pot_bulk_nxi}(b).

%%%%%%%%%%%%%%%%%%%%%
%%%%%%%%%%%%%%%%% 
%%%%%%%%%%%%%%%%%%%%%
\begin{figure}[t]
  \begin{center}
\mbox{\includegraphics[width = 0.48 \textwidth] {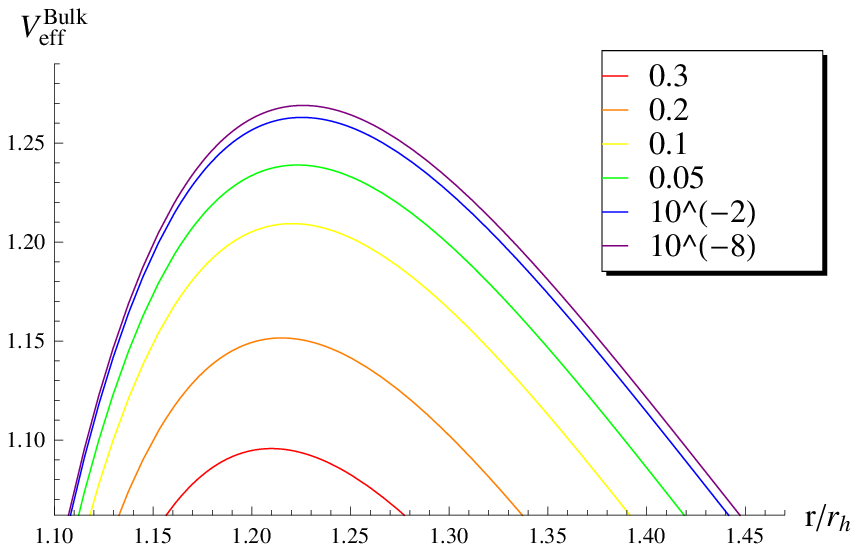}}
\hspace*{-0.1cm} {\includegraphics[width = 0.48 \textwidth]
{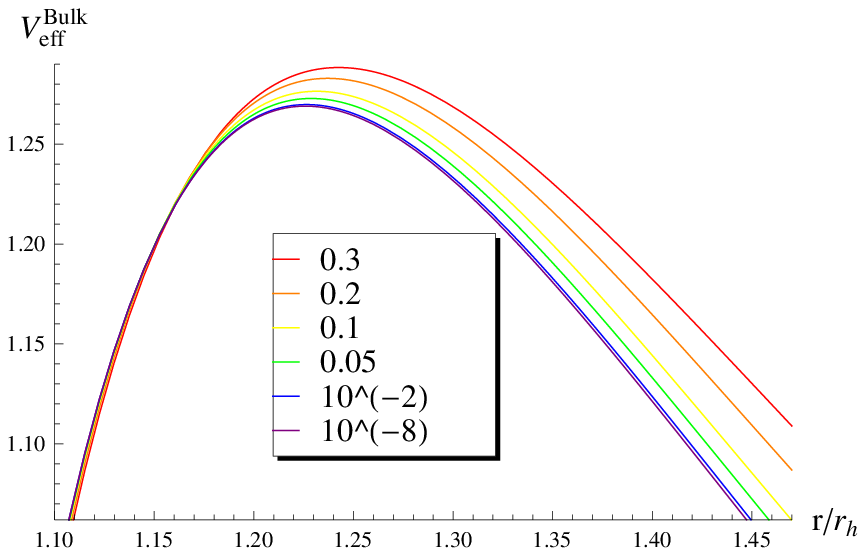}}
    \caption{Effective potential for bulk scalar fields for $l=0$, $n=2$,
    and variable $\Lambda=10^{-8}, 10^{-2},0.05,0.1,0.2,0.3$: 
    magnifications of the peak area for \textbf{(a)} $\xi=0$, and
    \textbf{(b)} $\xi=0.5$.}
   \label{pot_bulk_Lam}
  \end{center}
\end{figure}
%%%%%%%%%%%%%%%%%

The dependence of the bulk potential on the cosmological constant $\Lambda$
is similar to the one on the brane. Figures \ref{pot_bulk_Lam}(a,b) depict
the profile of the gravitational barrier for various values of $\Lambda$
and for minimal, $\xi=0$, and an arbitrary non-minimal coupling, $\xi=0.5$,
respectively.
Again, we find that for zero or small values of the coupling parameter
the height of the barrier decreases with the value of the cosmological
constant while beyond a critical value it starts to increase.

%%%%%%%%%%%%%%%%%%%%%%%%%%%%%%%%%%%

\subsection{The Analytical Solution} 

In order to solve analytically the scalar equation of motion in the bulk,
we follow the same approximation technique as in section 3.1. Near the
black-hole horizon, we make the same change of variable $r \rightarrow f$,
given in Eq. (\ref{newco-f}), in terms of which the radial equation
(\ref{radial_bulk}), at $r \simeq r_h$, becomes
%%%%%%%%%%%
\begin{eqnarray}
&~& \hspace*{-1.6cm} 
f\,(1-f)\,\frac{d^2 R}{df^2} + (1-B_h\,f)\,\frac{d R}{df}
+ \biggl[\,\frac{(\omega r_h)^2}{A_h^2}+ \frac{(\omega r_h)^2}{A_h^2 f} 
- \frac{\lambda_h\,(1-\tilde \Lambda r_h^2)}
{A_h^2 (1-f)}\,\biggr]R=0\,.
\label{NH-1-bulk}
\end{eqnarray}
%%%%%%%%%%
In the above, $A_h=(n+1)-(n+3)\,\tilde \Lambda r_h^2$ as before, whereas now
%%%%%%%%%
\beq
B_h \equiv 1+
\frac{4\tilde \Lambda r_h^2}{A_h^2}\,, \qquad \lambda_h=l(l+n+1)+\xi R_D r_h^2\,.
\eeq
%%%%%%%%%%%
Making the field redefinition $R(f)=f^{\alpha_1} (1-f)^{\beta_1} F(f)$, brings
the above again into the form (\ref{hyper}) of a hypergeometric equation.
The expressions of the $(a,b,c)$ parameters are now given by
%%%%%%%%%
\begin{eqnarray}
a_1&=&\alpha_1 + \beta_1 +\frac{1}{2}\,\left(B_h-1+
\sqrt{(B_h-1)^2-4 \omega^2 r_h^2/A_h^2}\right),\nonumber \\
b_1&=&\alpha_1 + \beta_1 + \frac{1}{2}\,\left(B_h-1-
\sqrt{(B_h-1)^2-4 \omega^2 r_h^2/A_h^2}\right), \\[1mm]
c_1&=&1+ 2 \alpha_1\,. \nonumber
\label{abc_bh_bulk}
\end{eqnarray}
%%%%%%%%%%
Using similar arguments as in Section 3, the power coefficients $\alpha_1$
and $\beta_1$ adopt exactly the same functional forms, i.e.
%%%%%%%%%%%%%%
\begin{equation}
\alpha_1 = - \frac{i \omega r_h}{A_h}\,,
\label{sol-a-bulk}
\end{equation}
%%%%%%%%%%
and 
%%%%%%%%%%%%
\begin{equation}
\beta_1 =\frac{1}{2}\,\biggl[\,(2-B_h) - \sqrt{(B_h-2)^2 + 
\frac{4\lambda_h\,(1-\tilde\Lambda r_h^2)}{A_h^2}}\,\,\biggr]\,.
\label{sol-be-bulk}
\end{equation}
%%%%%%%%%%%%%
with the only difference being the different forms of $B_h$ and $\lambda_h$. 
The general solution of the radial equation (\ref{NH-1-bulk}) in the bulk,
in the near-black-hole-horizon regime, again takes the form:
%%%%%%%%
\begin{equation}
R_{BH}(f)=A_1 f^{\alpha_1}\,(1-f)^{\beta_1}\,F(a_1,b_1,c_1;f)
\label{BH-gen-bulk}
\end{equation}
%%%%%%%%%

Near the cosmological horizon $r_c$, the radial equation (\ref{radial_bulk}),
in terms of the new variable $h(r) \simeq 1- \tilde \Lambda r^2$,
can be written as
%%%%%%%%%%%
\begin{eqnarray}
&& \hspace*{-1.5cm}h\,(1-h)\,\frac{d^2R}{dh^2} + 
\Bigl[1-\frac{(n+5)}{2}\,h\Bigr]\,\frac{dR}{dh} \nonumber \\[2mm]
&& \hspace*{2cm}+ \frac{1}{4}\,\biggl[\,\frac{(\omega r_c)^2}{h}- \frac{l(l+n+1)}{(1-h)}
-\xi (n+4)(n+3)\,\biggr] R=0\,.
\label{FF-1-bulk}
\end{eqnarray}
%%%%%%%%%%%%%%%
The field redefinition $R(h)=h^{\alpha_2} (1-h)^{\beta_2} X(h)$ brings the
above equation in the form of a hypergeometric equation with indices
%%%%%%%%%
\beq
a_2=\alpha_2 + \beta_2 +\frac{n+3}{4} +\frac{1}{4}\,\sqrt{(n+3)^2-4\xi (n+4)(n+3)}\,,
\eeq
%%%%%%%%%%%
\beq
b_2=\alpha_2 + \beta_2 + \frac{n+3}{4} -
\frac{1}{4}\,\sqrt{(n+3)^2-4\xi (n+4)(n+3)}\,, \qquad c_2=1+ 2 \alpha_2\,.
\eeq
%%%%%%%%%%
The power coefficients $\alpha_2$ and $\beta_2$ have now the form
%%%%%%%%%%%%%%
\begin{equation}
\alpha_2 = \frac{i \omega r_c}{2}\,, \qquad
\beta_2 =-\frac{(l+n+1)}{2}\,,
\label{alpha_beta-rc-bulk}
\end{equation}
%%%%%%%%%%%%%
and the general solution of the radial equation near the cosmological-horizon
is given again by Eq. (\ref{FF2-dS}).

The analysis after this point follows closely the one on the brane. The
two asymptotic solutions are shifted, then stretched and finally matched
at a intermediate radial regime. The identification of the multiplying
coefficients in front of the same powers of $r$, leads to the expressions
of the integration constants $B_1$ and $B_2$, and finally to the one of
the greybody factor that has again the form of Eq. (\ref{grey-brane-final}).

%%%%%%%%%%%%%%%%%%%%%%%%%%%%%%%%%%%%%%%%%%%%%%%%%%%%%%%%%%

\subsection{The Low-energy Limit} 

Similarly to the case of brane emission, we will now attempt to derive a
simplified expression for the greybody factor (\ref{grey-brane-final}) in
the bulk
in the low-energy limit, and study its asymptotic value both for minimal
and non-minimal coupling of the scalar field. The sets of $(a_i, b_i, c_i)$
bulk parameters near the black-hole and cosmological horizons, in the
limit of low energy, assume again the forms (\ref{abc_rh}) and (\ref{abc_rc}),
respectively, where now the $\omega$-independent quantities are defined as
%%%%%%%%%%%%%%%
\beq
\delta= \frac{1}{2}\left[B_h-\sqrt{(B_h-2)^2 +\frac{4 \lambda_h}{A_h^2}}\right],
\label{delta-bulk}
\eeq
%%%%%%%%%%%%
\beq
\epsilon= \frac{1}{2}\left[2-B_h-\sqrt{(B_h-2)^2 +\frac{4 \lambda_h}{A_h^2}}\right],
\label{epsilon-bulk}
\eeq
%%%%%%%
\beq
\eta_{\pm}= \frac{1}{4}\left[1-2l -n \pm \sqrt{(n+3)^2-4 \xi R_Dr_c^2}\right]\,.
\label{eta-bulk}
\eeq
%%%%%%%%
In the above, we have also taken the limit of small cosmological constant in order
to simplify further the analysis\footnote{In this limit, the quantity $B_h$ tends
to unity -- however, the limit of  $B_h \rightarrow 1$ will be taken after
the $\omega \rightarrow 0$ limit of the Gamma functions in order to keep the
two expansions distinct.}. The $\Sigma_i$ bulk
quantities are also identical to their brane analogues with the only difference
being that the exponent $l+1$ of $r_h$ and $r_c$ in the $\Sigma_1$, 
$\Sigma_3$ and $\Sigma_4$ quantities is replaced by $l+n+1$. 

In the case of minimal coupling, and for the lowest, dominant mode of
the scalar field, the aforementioned $\omega$-independent quantities 
are simplified to
%%%%%%%%%%%%
\beq
\delta=B_h-1\,, \qquad \epsilon=0\,, \qquad
\eta_\pm=\left[1,-\frac{(n+1)}{2}\right]\,. 
\eeq
%%%%%%%%%%%%%%%
Then, applying the low-energy limit, the $\Sigma_1$ quantity is simplified to
%%%%%%%%%%%%
\beq
\Sigma_1 \approx \frac{r_h^{n+1}\,\Gamma(1-2i\omega R_H)\,\Gamma(B_h-2)}
{\Gamma(B_h-1)\,\Gamma(-i \omega R_H)}
\rightarrow \frac{i \omega r_h^{n+2}}{(n+1)} + {\cal O} (\omega^2)\,.
\eeq
%%%%%%%%%%%%
Similarly, we obtain for the remaining $\Sigma_i$ quantities:
%%%%%%%%%%%%%%
\beq
\Sigma_2 \approx \Sigma_5 \approx \Sigma_6 \approx 1+ {\cal O} (\omega)\,,
\eeq
%%%%%%%%%%%%
\beq
\Sigma_3 \approx \frac{i \omega r_c^{n+2}}{(n+1)} + {\cal O} (\omega^2) \approx -\Sigma_4\,,
\eeq
%%%%%%%%%%%%%%
Under these approximations we finally obtain, from Eq. (\ref{grey-brane-final}),
the result
%%%%%%%%%%%%%%%%%%%%
\beq
|A^2|=1-\left|\frac{i \omega\,(r_c^{n+2}-r_h^{n+2})}{i \omega\,(r_c^{n+2}+
r_h^{n+2})}\right|^2
+{\cal O}(\omega) =
\frac{4 (r_hr_c)^{(n+2)}}{(r_c^{n+2}+r_h^{n+2})^2}+ {\cal O}(\omega)\,.
\label{geom_bulk}
\eeq
%%%%%%%%%%%%%%
Again, our general expression (\ref{grey-brane-final}), applied in the bulk,
correctly reproduces
the low-energy geometric value of the greybody factor for the mode $l=0$, in
accordance to previous higher-dimensional analyses \cite{KGB,Harmark}. Thus,
in the case of minimal coupling, scalar particles with very low energy have
a non-vanishing probability of being emitted by a higher-dimensional
Schwarzschild-de-Sitter spacetime, also in the bulk. However, due to the
presence of the number of extra dimensions $n$ in the exponents of $r_h$
and $r_c$ in the above expression, the magnitude of the low-energy 
asymptotic value in the bulk is significantly smaller compared to the one
on the brane.

Turning to the case of non-minimal coupling, we recall that our argument
of the vanishing of the greybody factor up to ${\cal O}(\omega)$ on the
brane did not use the explicit forms of the $(\delta, \epsilon, \eta_\pm)$
parameters. Since the functional forms of the $\Sigma_i$ quantities in
the bulk are identical to the ones they assumed on the brane, the same
argument holds in the case of bulk emission, too; according to this, all
partial waves, including the dominant one with $l=0$, do not assume
non-vanishing low-energy asymptotic values if $\xi$ is different from zero.
%%%%%%%%%%%%

The first non-vanishing term in the low-energy expansion of the greybody
factor, in the case where $\xi \neq 0$, is therefore of
${\cal O}(\omega^2)$. This has the explicit form
%%%%%%%%%%%%%%
\bea
|A|^2=  \frac{16 \pi^8 (r_h/r_c)^{l+n/2}\,(\omega r_h)^2 \sec^4(w\pi/2)
\left[\Gamma(\theta_{+})\,\Gamma(\theta_{-})\right]^{-2}}
{\sigma \sqrt{\sigma^2+4\xi R_D r_h^2}\left(\cos[\frac{\pi\sigma}{2}]+
\cos[\frac{\pi \sqrt{(n+3)^2-4\xi R_D r_c^2}}{2}]\right)^2\Gamma[\frac{\sigma}{2}]^2
\Gamma[w]^2\,\Gamma[\frac{1-w}{2}]^4}
\label{Crispino's generalization-bulk}
\eea
%%%%%%%%%%%%%%%
where $\sigma \equiv 2l+n+1$,
%%%%%%%%%%%%%
\beq
w=\frac{\sqrt{(2l+n+1)^2+4\xi R_D r_h^2}}{n+1}\
\eeq
%%%%%%%%
and
%%%%%%%%%%%
\beq
 \theta_{\pm}=\frac{1}{4} \left( 1-2 l-n \pm \sqrt{(n+3)^2-4 \xi R_D r_c^2}\right)\,.
\label{theta-bulk}
\eeq
%%%%%%%%%%%
The above reduces again to the 4-dimensional result derived in \cite{Crispino}
after setting $n=0$.

%%%%%%%%%%%%%%%%%%%%%%%%%%%%%%%%%%%%%%%%%%%%%%%%%%%%%%%%%%%%%%%%%%%%%%%%%

\subsection{Plotting our analytic result} 

\begin{figure}[t]
  \begin{center}
\mbox{\includegraphics[width = 0.42 \textwidth] {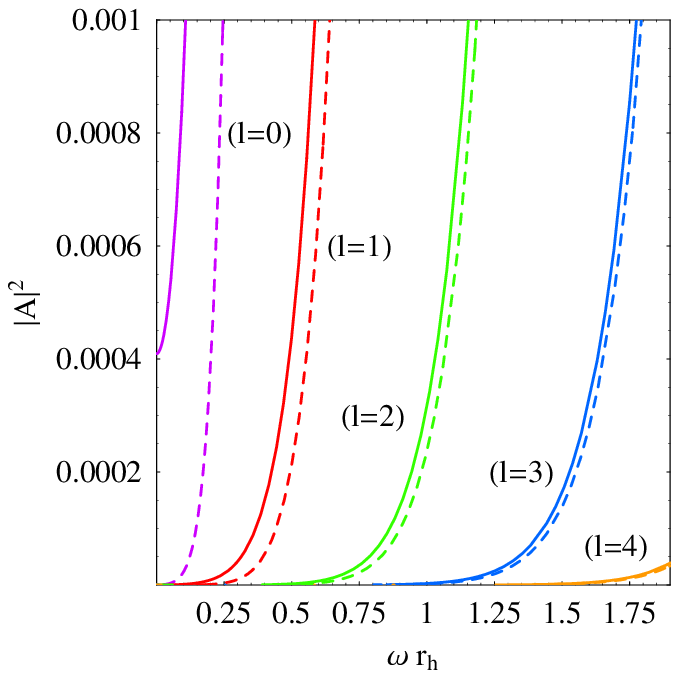}}
\hspace*{0.5cm} {\includegraphics[width = 0.42 \textwidth] {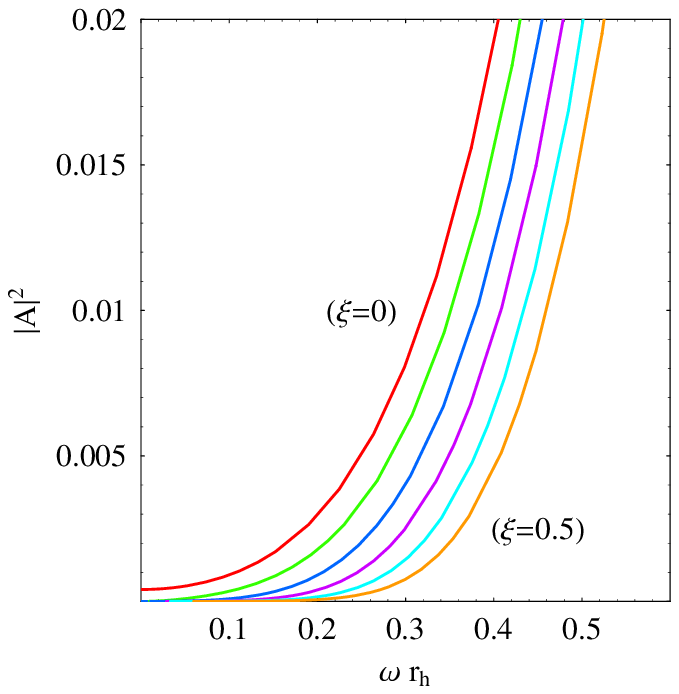}}
    \caption{ Greybody factors for bulk scalar fields, for $n=2$ and
	$\Lambda=0.1$, and: \textbf{(a)} for variable $l=0,1,2,3$ and $\xi=0$
	(solid lines) or $\xi=0.3$ (dashed lines); \textbf{(b)} for $l=0$ and
	variable $\xi=0,0.1,0.2,0.3,0.4,0.5$.}
   \label{grey_bulk_ellxi}
  \end{center}
\end{figure}

We will now study the dependence of the complete form of the greybody factor
for bulk scalar fields on both particle properties and spacetime properties. 
In Fig. \ref{grey_bulk_ellxi}(a), the greybody factor $|A|^2$ for the first
five partial waves with $l=0,1,2,3,4$, for $n=2$ and $\Lambda=0.1$ (again
in units of $r_h^{-2}$) is depicted; the suppression of the greybody factor
as the angular-momentum number $l$ increases is evident. The dashed lines
show again the same partial waves but for a non-vanishing value of the
non-minimal coupling parameter $\xi$: a suppression is again observed which
becomes less obvious as the angular-momentum number increases. In
Fig. \ref{grey_bulk_ellxi}(b), we focus again on the dominant partial
wave ($l=0$) and study in more detail the dependence of the greybody
factor on $\xi$: here, the suppression is significant and the transition
from the non-zero asymptotic value to a vanishing one in the low-energy limit
is again easily noticeable.

As it was noted in Section 4.1, the dependence of the effective potential for
bulk scalars on the non-minimal coupling parameter $\xi$ is much milder
compared to the case of brane scalars. In accordance to this, the modification
of the greybody factor as the value of $\xi$ changes is rather limited,
and this makes the use of zoom-in plots, like Figs. \ref{grey_bulk_ellxi}(a,b),
necessary. In addition, the presence of $n$ in the expression of
the low-energy asymptotic limit of $|A^2|$ for bulk emission (\ref{geom_bulk})
suppresses the latter significantly, therefore the magnification of the
low-energy regime is necessary in order to observe the transition from
the non-vanishing to the vanishing value of the greybody factor of the
lowest partial wave as the coupling parameter $\xi$ is turned on.

%%%%%%%%%%%%%%%%%%%%%
%%%%%%%%%%%%%%%%% 
%%%%%%%%%%%%%%%%%%%%
\begin{figure}[t]
  \begin{center}
\includegraphics[width = 0.4 \textwidth] {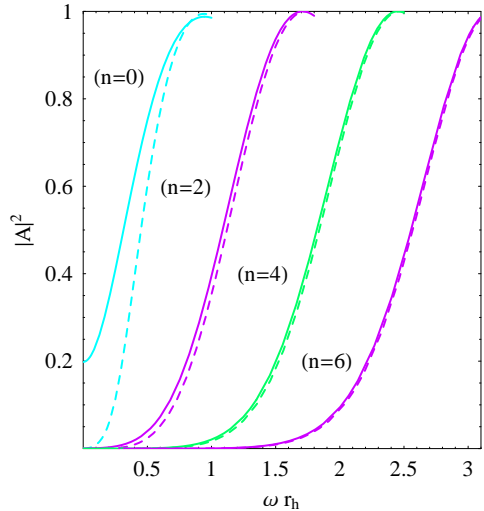}
    \caption{ Greybody factors for bulk scalar fields, for $l=0$, $\Lambda=0.1$
	and variable $n=0,2,4,6$ and for $\xi=0$ (solid lines) and $\xi=0.3$
	(dashed lines).}
   \label{grey_bulk_n}
  \end{center}
\end{figure}
%%%%%%%%%%%%%%%%%

The value of the greybody factor is suppressed also with the number of
extra spacelike dimensions, as shown in Fig. \ref{grey_bulk_n}. The
case of the lowest partial wave is again considered, and the dependence
on $n$ is shown both for minimal (solid lines) and non-minimal (dashed
lines) coupling. It is clear that the suppression with $\xi$ is more
significant for the smallest values of the parameter $n$. Note, that
due to poles appearing in the expressions of the Gamma functions,
that are present in the form of the greybody factor, no curves for
odd values of $n$ were possible to obtain.

%%%%%%%%%%%%%%%%%%%
\begin{figure}[b!]
  \begin{center}
  \begin{tabular}{cc}
\includegraphics[width = 0.42 \textwidth] {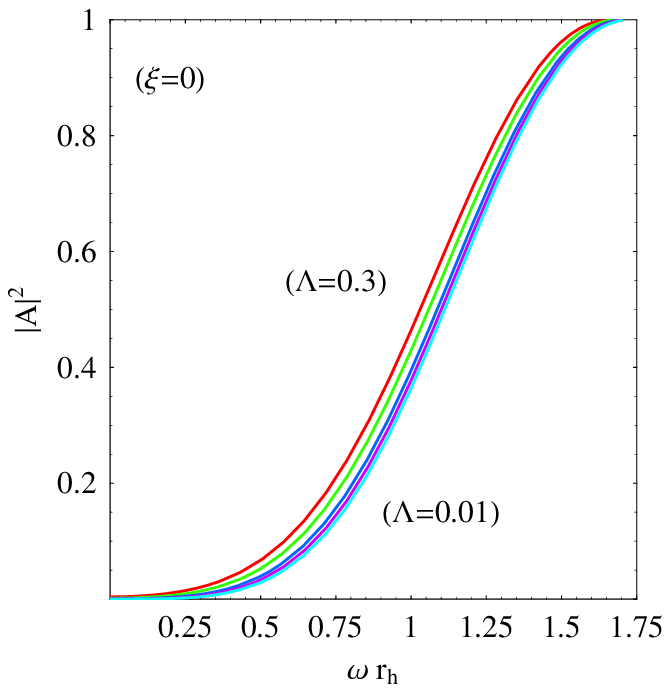}
& \hspace*{-0.1cm} {\includegraphics[width = 0.41 \textwidth]
{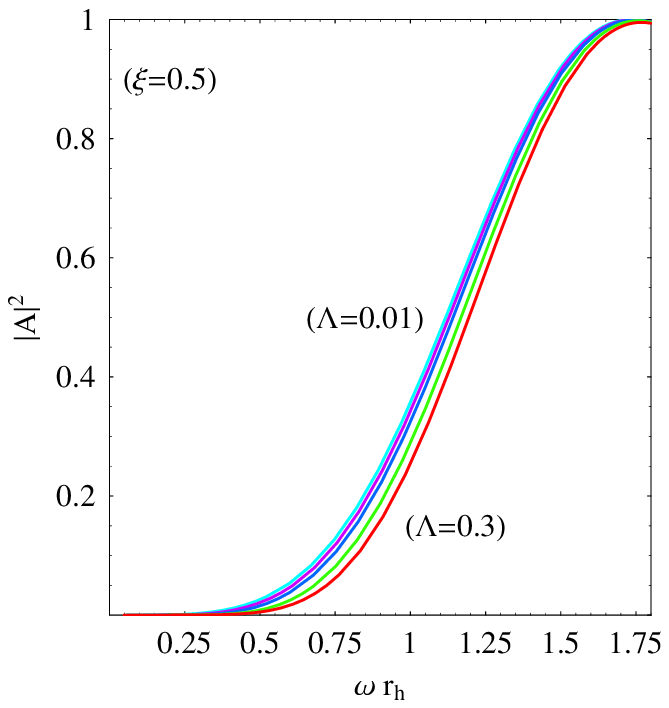}} \end{tabular}
    \caption{ Greybody factors for bulk scalar fields for $l=0$, $n=2$, and
	$\Lambda=$ 0.01, 0.05, 0.1, 0.2, 0.3, and \textbf{(a)} for $\xi=0$, and
	\textbf{(b)} $\xi=0.5$.}
   \label{grey_bulk_Lam}
  \end{center}
\end{figure}
%%%%%%%%%%%%%%%%%

In Figs. \ref{grey_bulk_Lam}(a,b), we finally depict the dependence of the
greybody factor on the bulk cosmological constant, for two values of
the non-minimal coupling parameter, $\xi=0$ and 0.5, and for the dominant
partial wave with $l=0$. In accordance to the behaviour of the effective
potential, we find that the value of $|A^2|$ is indeed enhanced with
the bulk cosmological constant provided that the value of $\xi$ is
small; beyond a certain value of the coupling parameter, the situation
is reversed, and the presence of $\Lambda$ hinders the emission of
scalar fields in the bulk.

%%%%%%%%%%%%%%%%%%%%%%%%%%%%%%%%%%%%%%%%%%%%%%%%%%%%%%%%%%%%%%%%%%%%5

\section{Conclusions}

In the context of the investigation of the Hawking radiation spectra
emitted by four- and higher-dimensional black holes, the greybody factors
of various species of fields for propagation in these spacetimes have been
intensively studied. For the case of a Schwarzschild-de-Sitter black holes,
the presence of the cosmological constant increased the complexity of the
field equations and made the analytic treatment particularly difficult.
Until recently, the existing analytic studies covered the case of only the
lowest partial mode of a scalar field in the low-energy approximation
\cite{KGB, Harmark}. The latest study \cite{Crispino} extended this
analysis to the case of arbitrary partial mode but in the strictly
four-dimensional case. 

In this work, we considered the emission of scalar particles by a
higher-dimensional Schwarzschild-de-Sitter black hole both on the
brane and in the bulk. Our analysis applied for arbitrary partial modes
and, in principle, is valid only at the low-energy regime;
however, especially in the case of brane emission where the matching
of the solutions did not impose a constraint on the energy of the emitted
particle, we envisage that our results may hold well beyond the 
low-energy regime. In addition, a particular effort was made to take
into account the effect of the bulk cosmological constant both close
and far away from the black-hole horizon. Near the black-hole horizon
particularly, no simplification of the metric tensor was adopted -- in
contrast to previous studies -- and an appropriately chosen new radial
coordinate was employed that allowed us to analytically integrate the
scalar radial equation in an exact Schwarzschild-de-Sitter spacetime. 

Starting from the more phenomenologically interesting case of the
emission of scalar fields on the brane, we first derived the analytic
form of the greybody factor by following the matching technique of the
two asymptotic solutions in the black-hole and cosmological horizons.
This expression was first studied analytically, with its low-energy
limit computed both for the cases of minimal and non-minimal couplings. 
In the first case, it was demonstrated that our general expression,
for the dominant lowest mode, indeed reduced to the non-vanishing
low-energy asymptotic limit found in previous works. We analytically
showed that in the presence of a non-minimal coupling of the scalar
field with the induced-on-the-brane Ricci scalar, this asymptotic
limit vanishes both at ${\cal O}(\omega^0)$ and ${\cal O}(\omega)$;
the analytic expression of the first non-vanishing term at
${\cal O}(\omega^2)$ was computed and
presented. Then, we studied the profile of the complete result
for the greybody factor in terms of particle and spacetime properties.
Our results showed that the greybody factor, and thus the emission of
scalar particles on the brane by a Schwarzschild-de-Sitter black hole,
is suppressed as either the angular-momentum number or the non-minimal
coupling parameter increases. The same suppression is observed with
the number of extra spacelike dimensions that exist transverse to the
brane. In terms of the bulk cosmological constant, we finally demonstrated
that its presence acts in two contradictory ways: as a homogeneously
distributed energy over the whole spacetime, it subsidizes the energy of
the emitted particle and increases its greybody factor; but, as an
effective mass term for the scalar field through the non-minimal
coupling term, it decreases its greybody factor -- the net effect
depends crucially on the value of the non-minimal coupling parameter.  

The emission of scalar fields in the bulk by the higher-dimensional
Schwarzschild-de-Sitter spacetime was studied next. The corresponding
analysis led again to a general expression for the greybody factor
valid for arbitrary partial modes and values of the non-minimal coupling
parameter with the latter defining the coupling of the bulk scalar field
to the curvature of the complete higher-dimensional spacetime. For the
case of minimal coupling, the low-energy asymptotic value for the
lowest partial mode was again computed and shown to agree with previous
studies. A similar analysis for the case of non-minimal coupling,
to the one presented for the brane emission, was also performed. 
The dependence of its complete expression on the various parameters
of the theory was also studied in detail and found to follow the
same pattern as in the case of brane emission, the main characteristic
being the milder dependence of the greybody factor on the non-minimal
coupling parameter. 

Our results are in excellent agreement with previous analytic results
in higher-dimensional \cite{KGB} and four-dimensional \cite{Crispino}
contexts and also follow closely the exact results produced by 
numerical analysis in the same works. Although, for the case of
minimal coupling, the study of the corresponding Hawking radiation
spectra by a higher-dimensional Schwarzschild-de-Sitter black hole 
has been performed \cite{KGB}, the one for a non-minimal coupling, 
in the same higher-dimensional context,
is still lacking. As we believe that the dual role of the cosmological
constant in that case may significantly affect the radiation spectra
and, perhaps, the bulk-to-brane ratio of emitted energy, we hope to
soon report results from such an analysis \cite{KPP2}. 

%%%%%%%%%%%%%%%%%%%%%%%%%%%%%%%%%%%%%%%%%%%%%%%%%%%%%%%%%%%%%%%%%%%%%%

{\bf Acknowledgements.} This research has been co-financed by the European
Union (European Social Fund - ESF) and Greek national funds through the
Operational Program ``Education and Lifelong Learning'' of the National
Strategic Reference Framework (NSRF) - Research Funding Program: 
``ARISTEIA. Investing in the society of knowledge through the European
Social Fund''. Part of this work was supported by the COST Actions MP0905
``Black Holes in a Violent Universe'' and MP1210 ``The String Theory Universe''.

%%%%%%%%%%%%%%%%%%%%%%%%%%%%%%%%%%%%%%%%%%%%%%%%%%%%

\end{document}